\documentclass[sigconf, authorversion]{acmart}
\usepackage{subcaption}
\usepackage{bbding}
\usepackage{multirow}
\usepackage[toc,page]{appendix}
\AtBeginDocument{%
  }

\copyrightyear{2022} 

\acmYear{2022} 
\setcopyright{acmcopyright}\acmConference[MM '22]{Proceedings of the 30th ACM International Conference on Multimedia}{October 10--14, 2022}{Lisboa, Portugal}
\acmBooktitle{Proceedings of the 30th ACM International Conference on Multimedia (MM '22), October 10--14, 2022, Lisboa, Portugal}
\acmPrice{15.00}
\acmDOI{10.1145/3503161.3547845}
\acmISBN{978-1-4503-9203-7/22/10}

\begin{document}

\title[Hybrid Spatial-Temporal Entropy Modelling for Neural Video Compression]{Hybrid Spatial-Temporal Entropy Modelling for \\ Neural Video Compression}

\author{Jiahao Li}
\affiliation{%
	\institution{Microsoft Research Asia}
	\city{Beijing}
	\country{China}}
\email{li.jiahao@microsoft.com}
 
 \author{Bin Li}
 \affiliation{%
 	\institution{Microsoft Research Asia}
 	\city{Beijing}
 	\country{China}}
 \email{libin@microsoft.com}
 
 \author{Yan Lu}
 \affiliation{%
 	\institution{Microsoft Research Asia}
 	\city{Beijing}
 	\country{China}}
 \email{yanlu@microsoft.com}

\renewcommand{\shortauthors}{Li et al.}

\begin{abstract}
For neural video codec, it is critical, yet challenging, to design an efficient entropy model which can accurately predict the probability distribution of the quantized latent representation. However, most existing video codecs directly use the ready-made entropy model from  image codec to encode the residual or motion, and do not fully leverage the spatial-temporal characteristics in video. To this end, this paper proposes a powerful entropy model which efficiently captures both spatial and temporal dependencies. In particular, we introduce the latent prior which exploits the correlation among the latent representation to squeeze the temporal redundancy. Meanwhile, the dual spatial prior is proposed to reduce the spatial redundancy in a parallel-friendly manner. In addition, our entropy model is also versatile. Besides estimating the probability distribution, our entropy model also generates the quantization step at spatial-channel-wise. This content-adaptive quantization mechanism not only helps our codec achieve the smooth rate adjustment in single model but also improves the final rate-distortion performance by dynamic bit allocation. Experimental results show that, powered by the proposed entropy model, our neural codec can achieve 18.2\% bitrate saving on UVG dataset when compared with H.266 (VTM) using the highest compression ratio configuration.  It makes a new milestone in the development of neural video codec. The codes are at \url{https://github.com/microsoft/DCVC}.
\end{abstract}

 

 
\maketitle
 
\section{Introduction}
\label{sec_introduction}
 Recent years have witnessed the flourish of neural image codec. During the  development, many works focus on the design of entropy model to accurately predict the  probability distribution  of the quantized latent representation, like factorized model \cite{ball2017endtoend}, hyper prior \cite{balle2018variational}, auto-regressive prior \cite{minnen2018joint}, mixture Gaussian model \cite{cheng2020learned},  transformer-based model \cite{koyuncu2022contextformer}, and so on. Benefited from these continuously improved entropy models, the compression ratio of neural image codec has outperformed the best traditional codec H.266 intra coding \cite{bross2021overview}.  Inspired by the success of neural image codec, recently the neural video codec  attracts
more and more attentions.  

Most existing works on  neural video codec  can be roughly classified into three categories: residual coding-based, conditional coding-based, and 3D autoencoder-based solutions. Among them, many methods \cite{liu2020learned,Rippel_2021_ICCV,hu2020improving,lu2020content,lu2020end,lin2020m,hu2021fvc,agustsson2020scale,rippel2019learned,Djelouah_2019_ICCV,yang2021learning,wu2018video,liu2020neural} belong to the residual coding-based solution. The residual coding comes from the traditional hybrid video codec. Specifically, the motion-compensated prediction is first generated, and then its residual  with the current frame is coded.  
For conditional coding-based solutions \cite{ladune2021conditional,ladune2021conditional222,ladune2020optical,li2021deep}, the temporal frame or feature is served as condition for the coding of the current frame. When compared with residual coding,  conditional coding  has lower or equal entropy bound \cite{ladune2020optical}. As for 3D autoencoder-based solution \cite{pessoa2020end, habibian2019video, sun2020high}, it is a natural extension of neural image codec by expanding the input dimension. But it  brings larger encoding delay and significantly increases the memory cost.  
In a summary, most these existing works  focus on how to generate the optimized latent representation by exploring different data flows or network structures. As for the entropy model, they usually directly use the ready-made solutions (e.g., hyper prior \cite{balle2018variational} and auto-regressive prior \cite{minnen2018joint}) from neural image codec to code the  latent representation. The spatial-temporal correlation has not been fully explored in the design of entropy model for video. Thus, the RD (rate-distortion) performance of previous SOTA (state-of-the-art)  neural video codec \cite{sheng2021temporal} is limited and  only 
slightly better than H.265, which was released in 2013.
   
Therefore, this paper proposes a comprehensive entropy model which can efficiently leveraging both spatial and temporal correlations, and then helps the neural video codec outperform the latest traditional  standard H.266. 
In particular, we  introduce the latent prior and dual spatial prior.
The latent prior explores the temporal correlation of the latent representation across frames. The quantized latent representation of the previous frame is used to predict the  distribution of that in the current frame.
Via the cascaded training strategy, the propagation chain  of latent representation is formed. It enables us to build the implicit  connection between the latent representation of the current frame and that of the long-range reference frame. Such connection helps the neural codec further squeeze the temporal redundancy among the latent representation.

In our entropy model, the dual spatial prior is proposed to reduce the spatial redundancy. Most existing neural codecs rely on the auto-regressive prior \cite{minnen2018joint} to explore the spatial correlation. However,  auto-regressive prior is a serialized solution and follows a strict scaning order. Such kind of solution is parallel-unfriendly and inferences in a very slow speed. By contrast, our dual spatial prior is a two-step coding solution following the checkerboard 
context model \cite{he2021checkerboard}, which is much more time-efficient.  
In \cite{he2021checkerboard},  all channels use the same coding order (even positions are always first coded and then are used as context for the  odd positions). It cannot efficiently cope with various video contents because sometimes coding the even positions first has worse  RD performance than coding the odd positions first. Thus, to solve this problem, our dual spatial prior introduces the mechanism that first codes the half latent representation of both odd and even  positions, and then the coding of the left latent representation can benefit from the contexts from  all positions. 
At the same time, the correlation across the channel is also exploited during the two-step coding.
Without bringing extra coding dependency, out dual spatial prior  makes the scope of spatial context doubled and exploits the channel context. These will result in more accurate prediction on  distribution.
 
 
For neural video codec, another challenge is how to achieve smooth rate adjustment in single model. For traditional codec, it is achieved by adjusting the quantization parameter. 
However, most neural codecs lack such capability and use fixed quantization step (QS). To achieve different rates, the codec needs to be retrained. It brings huge training and model storage burden. To solve this problem, we introduce an adaptive quantization mechanism at multi-granularity levels, which is powered by our entropy model. In our design, the whole QS is  determined 
at three different granularities. First, the global QS is set by the user for the specific target rate. Then it is multiplied  by the channel-wise QS because different channels contain information with different importance, similar to the channel attention mechanism \cite{hu2018squeeze}. At last, the spatial-channel-wise  QS generated by our entropy model is multiplied. 
This can help our codec cope with various video contents and achieve precise rate adjustment at each position.
 In addition, it is noted that  using entropy model to learn the  QS  not only helps our codec  obtain the capability of smooth rate adjustment in single model but also improves the final RD performance. This is because the entropy model will learn to allocate more bits to the more important contents which are vital for the reconstruction of the current and following frames. This kind of content-adaptive quantization mechanism enables the dynamic bit allocation  to boost the final compression ratio.

Powered by our versatile entropy model, our neural codec with only single model can achieve significant bitrate saving over previous SOTA neural video codecs. For example, there is a significant 57.1\%  bitrate saving over   DCVC \cite{li2021deep}  on UVG dataset \cite{uvg}. Better yet, our neural codec has outperformed the best traditional codec H.266 (VTM) \cite{bross2021overview,VTM}, which uses  the low delay configuration with the highest compression ratio setting. For UVG dataset, an average of 18.2\% bitrate saving is achieved over  H.266 (VTM), when oriented to PSNR. If oriented to MS-SSIM, the corresponding bitrate saving is 35.1\%.
These substantial improvements show that our model makes a new milestone in the development of neural video codec.

Our contributions are summarized as follows:
\begin{itemize}
    \item For neural video codec, we design  a powerful and parallel-friendly entropy model to improve the prediction on probability distribution. The proposed latent prior and dual spatial prior can efficiently capture the temporal and spatial dependency, respectively. This helps us further squeeze the  redundancy in video.
    \item Our entropy model is also versatile. Besides the distribution parameters, it also generates the QS at spatial-channel-wise. Via the multi-granularity quantization mechanism,  our neural codec is able to achieve smooth rate adjustment in single model. Meanwhile, the QS at spatial-channel-wise is content-adaptive and   improves the final RD performance by dynamic bit allocation.
    \item Powered by the proposed entropy model, our neural video codec pushes the compression ratio to a new height.  To the best of our knowledge, our codec is the first end-to-end neural video codec to exceed   H.266 (VTM) using the highest compression ratio configuration.
    The bitrate saving over H.266 (VTM) is  18.2\% on UVG dataset in terms of PSNR. If oriented to MS-SSIM, the bitrate saving is even higher. 
\end{itemize}

\section{Related Work}
\label{sec_related_work}
\subsection{Neural Image Compression}
The neural image codec has developed rapidly in recent years. The early work \cite{theis2017lossy} uses a compressive autoencoder-based framework to achieve similar RD performance  with  JPEG 2000. Recently, many works focus on the design of entropy model. 
Ball{\'e} \textit{ et al.} \cite{ball2017endtoend} proposed the factorized model and got better RD performance than JPEG 2000. The  hyper prior \cite{balle2018variational} introduces the hierarchical design and uses additional bits to estimate the distribution, which gets comparable results with H.265. Subsequently, the auto-regressive prior \cite{minnen2018joint} was proposed to explore the spatial correlation. It obtains higher compression ratio together with  hyper prior. However, auto-regressive prior inferences in a serialized order for all spatial positions, and thus it is quite slow. 
The checkerboard context model \cite{he2021checkerboard} solves the complexity problem by introducing the two-step coding. In addition, to further boost the compression ratio, the mixture Gaussian model \cite{cheng2020learned} was proposed and its RD results are on par with H.266. Recently, the vision transformer attracts lots of attention, and the corresponding  entropy model \cite{ koyuncu2022contextformer}  helps the neural image codec  exceed H.266 intra coding.

\begin{figure*}
		\begin{center}
			\includegraphics[width=\linewidth]{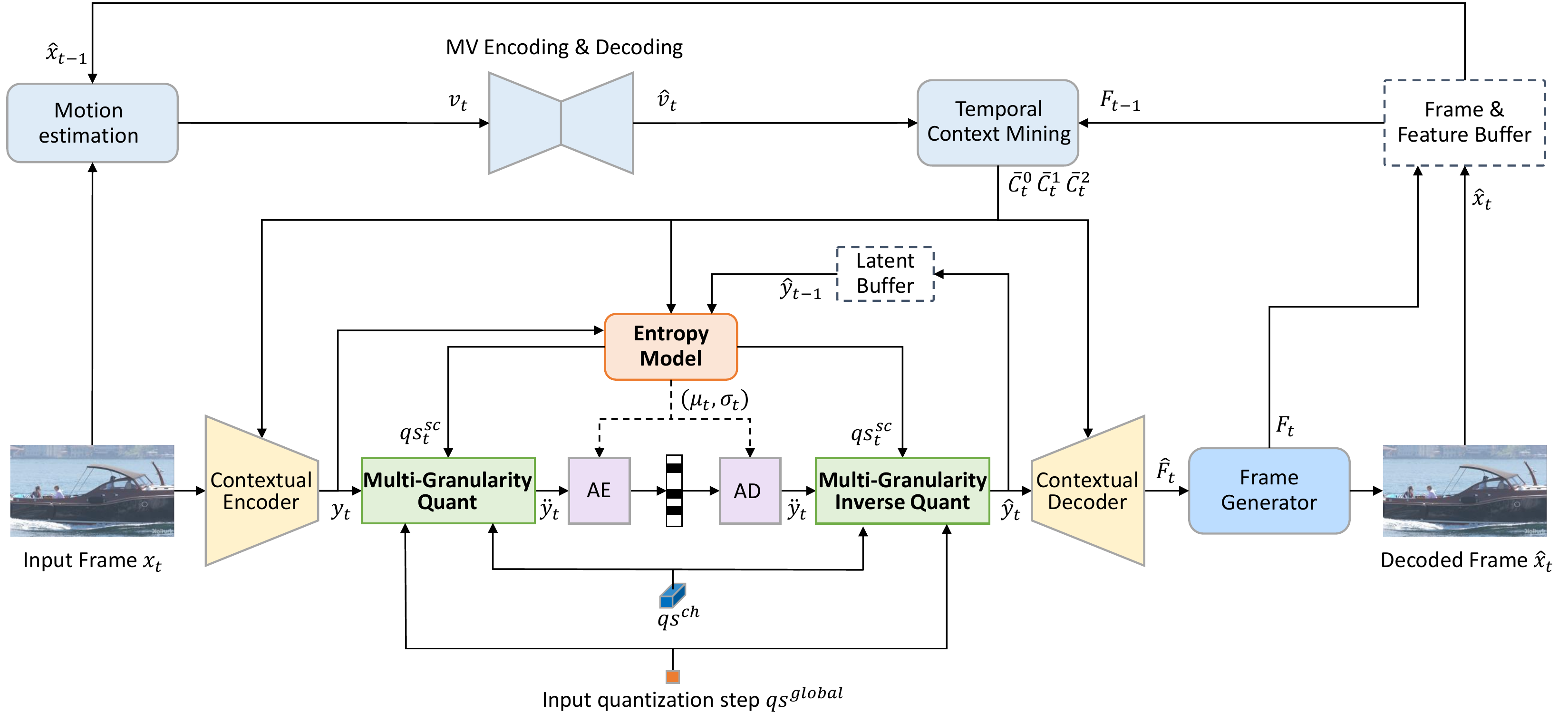}
		\end{center}
		\vspace{-0.3cm}
		\caption{The overall framework of our method.  Quant shorts for quantization.  AE and AD are arithmetic encoder and decoder. 
		The entropy model and  quantization mechanism for coding motion vector follow the similar design with those of $y_t$, and we omit them for simplification.  
		} 
		\label{fig_framework}
\end{figure*}
\subsection{Neural Video Compression}
The success of neural image codec also pushes the development of neural video codec. The pioneering work DVC \cite{lu2019dvc} follows the traditional codec, and uses the residual coding-based framework where  motion-compensated prediction is first generated and then the residual is coded using hyper prior \cite{balle2018variational}. With the help of auto-regressive prior \cite{minnen2018joint}, its following work DVCPro achieves higher compression ratio. 

Most recent works follow this motion estimation \cite{ranjan2017optical,hui2020lightweight} and residual coding-based framework. Then more advanced networks structures are proposed for generating the optimized residual or motion. For example, the residual is adaptively scaled by learned parameter in \cite{yang2020hierarchical}. The optical flow estimation in scale space \cite{agustsson2020scale} was proposed to reduce the residual energy in fast motion area. In \cite{hu2020improving}, the rate distortion optimization is applied to improve the 
coding of motion. The deformable compensation \cite{hu2021fvc} is used to improve the prediction in feature space.  Lin \textit{et al.} \cite{lin2020m} proposed using multiple reference frames to reduce the residual energy. In \cite{lin2020m,Rippel_2021_ICCV}, the motion prediction    is introduced to improve the coding efficiency of motion. 

Besides the residual coding, other coding frameworks are also investigated. For example,    3D autoencoder \cite{pessoa2020end, habibian2019video, sun2020high} was proposed to encode the multiple frames simultaneously. It is a natural extension of neural image codec by expanding the input dimension. However, such kind of framework will bring significant encoding delay and is not suitable for real-time scenarios. Another emerging coding framework is the conditional coding whose entropy bound is lower  than or equal to residual coding \cite{ladune2020optical}.  For example, Ladune \textit{et al.} \cite{ladune2021conditional,ladune2021conditional222,ladune2020optical} 
used the conditional coding to code the foreground contents. In DCVC \cite{li2021deep}, the condition is the extensible high-dimension feature rather than the 3-dimension predicted frame. The following work \cite{sheng2021temporal} further boosts the compression ratio by introducing the feature propagation and multi-scale temporal contexts.

However, most existing neural video codecs  focus on  how to generate the optimized latent representation and design the network structures therein. As for the entropy model used for coding the latent representation, the ready-made solutions from neural image codec are directly used. In this paper, we focus on the entropy model design by efficiently leveraging both spatial and temporal correlations. Actually, some works also have begun to investigate it. For example, the conditional entropy coding was proposed in \cite{liu2020conditional}. 
The temporal context prior extracted from temporal feature is used in \cite{li2021deep, sheng2021temporal}. Yang \textit{et al.} \cite{yang2021learning}  proposed the recurrent entropy model. However, these works \cite{liu2020conditional, li2021deep, sheng2021temporal, yang2021learning}  focus more on utilizing temporal correlation. Although the work in \cite{li2021deep} also investigates the spatial correlation, the auto-regressive prior is used and leads to a very slow coding speed. An entropy model which not only fully exploits the spatial-temporal correlation but also has low complexity is desired. To meet this requirement, we specially design the latent prior and time-efficient dual spatial prior to equip the entropy model, and push the compression ratio to a new height. In addition, all these methods  \cite{liu2020conditional, li2021deep, sheng2021temporal, yang2021learning} need to  train separate model for each rate point. By contrast, we design a multi-granularity quantization, which is powered by our entropy model. This content-adaptive quantization mechanism helps our codec achieve smooth rate adjustment in single model. The  rate adjustment in single model was investigated for neural image codec via the gain unit \cite{cui2021asymmetric}. And yet, 
to the best of our knowledge, our codec   is the first  neural video codec to obtain such capability via our multi-granularity quantization.

\begin{figure*}
		\begin{center}
			\includegraphics[width=\linewidth]{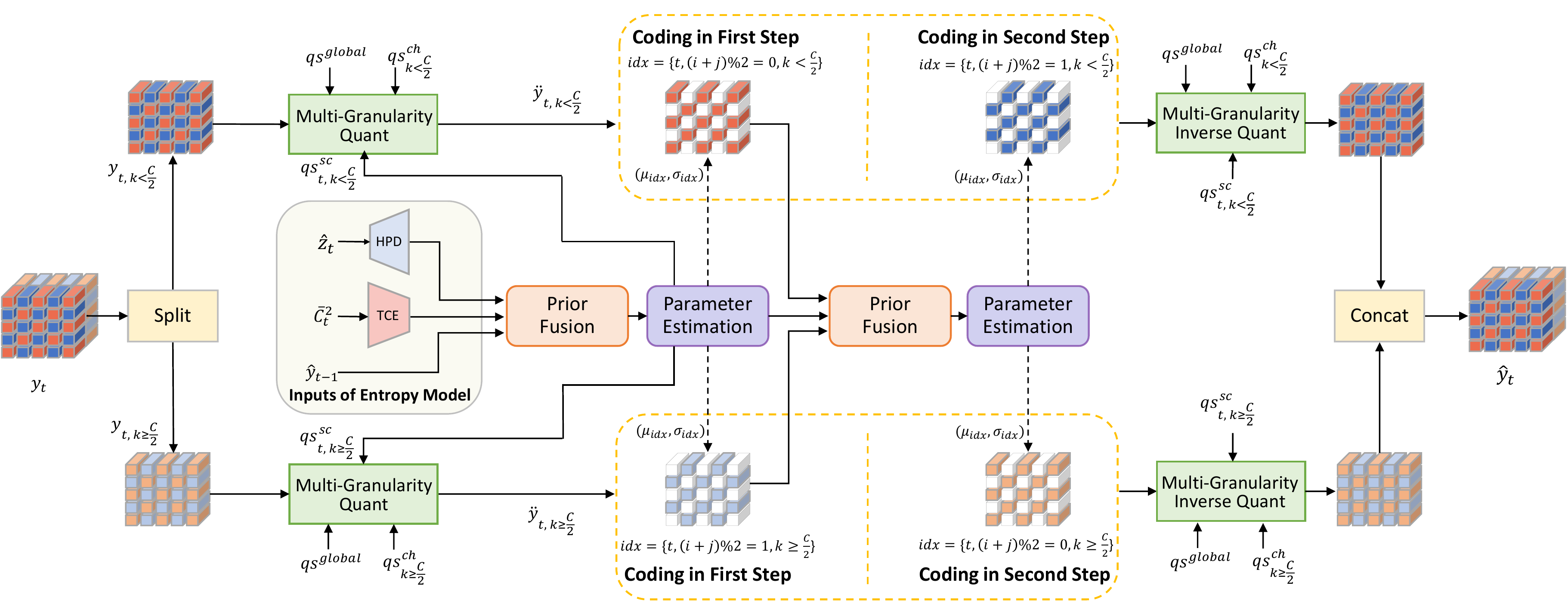}
		\end{center}
		\vspace{-0.4cm}
		\caption{The illustration of our entropy model and the coding of quantized latent representation. 
		$i$, $j$, and $k$ indicate the height, width, and channel indexes, respectively. $C$ is the channel number. $qs^{global}$, $qs^{ch}$, and $qs^{sc}$ are global, channel-wise, spatial-channel-wise quantization steps, respectively. HPD shorts for hyper prior decoder. TCE means temporal context encoder. For simplification, the arithmetic encoder and decoder following the parameter estimation are omitted in the two-step coding.
		} 
		\label{fig_entropy_model}
\end{figure*}
\section{Proposed Method}
	
\subsection{Framework Overview} \label{sub_sec_proposed_overview}
To achieve  higher compression ratio for the  input frame $x_t$ ($t$ is the frame index), we  adopt the conditional coding-based framework rather than the residual coding-based framework. Specifically, we follow the DCVC \cite{li2021deep} and its improved work \cite{sheng2021temporal}, and then redesign the core modules therein. The framework of our codec is presented in Fig.  \ref{fig_framework}. As shown in this figure, the entire coding process can be roughly divided into three steps: temporal context generation, contextual encoding/decoding, and reconstruction.

\textbf{Temporal context generation.} 
To fully explore the temporal correlation, we generate the multi-scale contexts {$\bar{C}^{0}_{t}$, $\bar{C}^{1}_{t}$,$\bar{C}^{2}_{t}$} at different  resolutions via the temporal context mining module (more details of this module can be found in \cite{sheng2021temporal}). 
To carry  richer information, we also use the temporal feature $F_{t-1}$ rather than  previous decoded frame  $\hat{x}_{t-1}$ as the module input. 
As  for motion estimation, we adopt the light-weight SPyNet \cite{ranjan2017optical} for acceleration.  

\textbf{Contextual encoding/decoding.} Conditioned by the multi-scale contexts, the current frame $x_t$ is transformed into latent representation $y_t$ by the contextual encoder. To achieve bitrate saving, the $y_t$ is quantized to $\ddot{y}_t$  before being sent to the arithmetic encoder which  generates the bit-stream. 
During the decoding, $\ddot{y}_t$  is decoded from  bit-steam by  arithmetic decoder  and inversely quantized to  $\hat{y}_{t}$. Also conditioned on the multi-scale contexts, the contextual decoder decodes the high-resolution feature $\hat{F}_t$  from $\hat{y}_{t}$. 
In this encoding/decoding process, how to accurately estimate the  distribution of $\ddot{y}_t$ by the entropy model is vital for  bitrate reduction. To this end, we propose the hybrid spatial-temporal entropy model (Section \ref{sub_sec_entropy_model}). To support smooth  rate adjustment in single model, the multi-granularity quantization powered by our entropy model is proposed (Section \ref{sub_sec_multirate}).

\textbf{Reconstruction.} After obtaining the high-resolution feature $\hat{F}_t$, our target is generating the high-quality reconstructed frame $\hat{x}_{t}$ via the frame generator. Different from DCVC \cite{li2021deep} and \cite{sheng2021temporal} only using plain residual blocks \cite{he2016deep}, we proposing using the W-Net \cite{xia2017w} based  structure which ties two U-Nets. Such kind of network design can effectively enlarge the receptive field of model with acceptable complexity. This results in stronger   generation ability for model. 

\subsection{Hybrid Spatial-Temporal Entropy Model}
\label{sub_sec_entropy_model}
For arithmetic coding, it needs to know the probability mass function (PMF) of $\ddot{y}_t$ to code it. However, we do not know its true PMF $p(\ddot{y}_t)$ and usually approximate it with an estimated PMF $q(\ddot{y}_t)$. The  cross-entropy $\mathbb{E}_{\ddot{y}_{t} \sim p}[-log_{2}  q(\ddot{y}_{t})]$ captures the  average number of bits needed by the arithmetic coding without considering the negligible overhead.
In this paper, we follow the existing work \cite{PyTorchVideoCompression} and assume that $q(\ddot{y}_{t})$ follows the Laplace distribution.   Thus, our target is  designing an  entropy model which can accurately estimate the distribution parameter of $q(\ddot{y}_{t})$ to reduce the cross-entropy.

To improve the estimation, for each element $\ddot{y}_{t,i,j,k}$ ($i$, $j$, and $k$ are the height, width, and channel indexes) therein, we need to fully mine the correlation between it and its  known information, e.g., the previous decoded latent representation, temporal context feature, and so on.
Theoretically, 
 $\ddot{y}_{t,i,j,k}$ may correlate to all these information from all previous decoded positions. For traditional codec, it is unable to explicitly exploit such correlation due to the huge space. Thus, traditional codec usually uses simple handcrafted rules to use context from  a few neighbour positions. By contrast, the deep learning enables the capability of automatically mining the correlation in huge space. 
 
 Thus, we propose  feeding manifold inputs to the entropy model, and  let the model extract the  complementary information from the rich high-dimensional inputs. The illustration of our   entropy model is shown in Fig. \ref{fig_entropy_model}. As shown in this figure, the inputs not only contain    the commonly-used hyper prior  $\hat{z}_{t}$ and  the temporal context  $\bar{C}^{2}_{t}$, but also include the  $\hat{y}_{t-1}$ as the latent prior. Although both $\bar{C}^{2}_{t}$ and $\hat{y}_{t-1}$ are from temporal direction, they have different characteristics. For example, $\bar{C}^{2}_{t}$ at 4x down-sampled resolution usually contains lots of motion information (more details can be found in \cite{sheng2021temporal}). By contrast, $\hat{y}_{t-1}$ is in the latent representation domain at 16x down-sampled resolution, and has more similar characteristics  with $y_t$. Thus, $\bar{C}^{2}_{t}$ and $\hat{y}_{t-1}$ can provide the  complementary auxiliary information to improve estimation. In addition, it is noted that we adopt the cascaded training strategy \cite{chan2021basicvsr, sheng2021temporal}, where the gradients will back propagate to multiple frames. Under such  training strategy, the propagation chain of latent representation is formed. It means that the connection between the latent representation of the current  frame and that of long-range reference frame is also built. Such connection is very helpful for extracting the correlation across the latent representation of multiple frames, and then results in more accurate  prediction on distribution.

Besides using  the latent prior $\hat{y}_{t-1}$ to enrich the input, our entropy model also adopts the dual spatial prior to exploit the spatial correlation. 
To pursue a time-efficient mechanism, we follow the checkerboard  model \cite{he2021checkerboard} rather than the commonly-used   auto-regressive model \cite{minnen2018joint} which seriously slows down the coding speed. However, all channels in the original checkerboard model use the same  coding order, namely the even positions are always first coded and then used as context for the coding of odd positions. Such coding order cannot handle various videos because sometimes coding the even positions first has worse RD performance than coding the odd positions first. Therefore, we design the dual spatial prior, where half channels in both even and odd  positions are first coded at the same time. 
\label{sub_sec_multirate}
\begin{figure}[t]
  \centering
  \includegraphics[width=\linewidth]{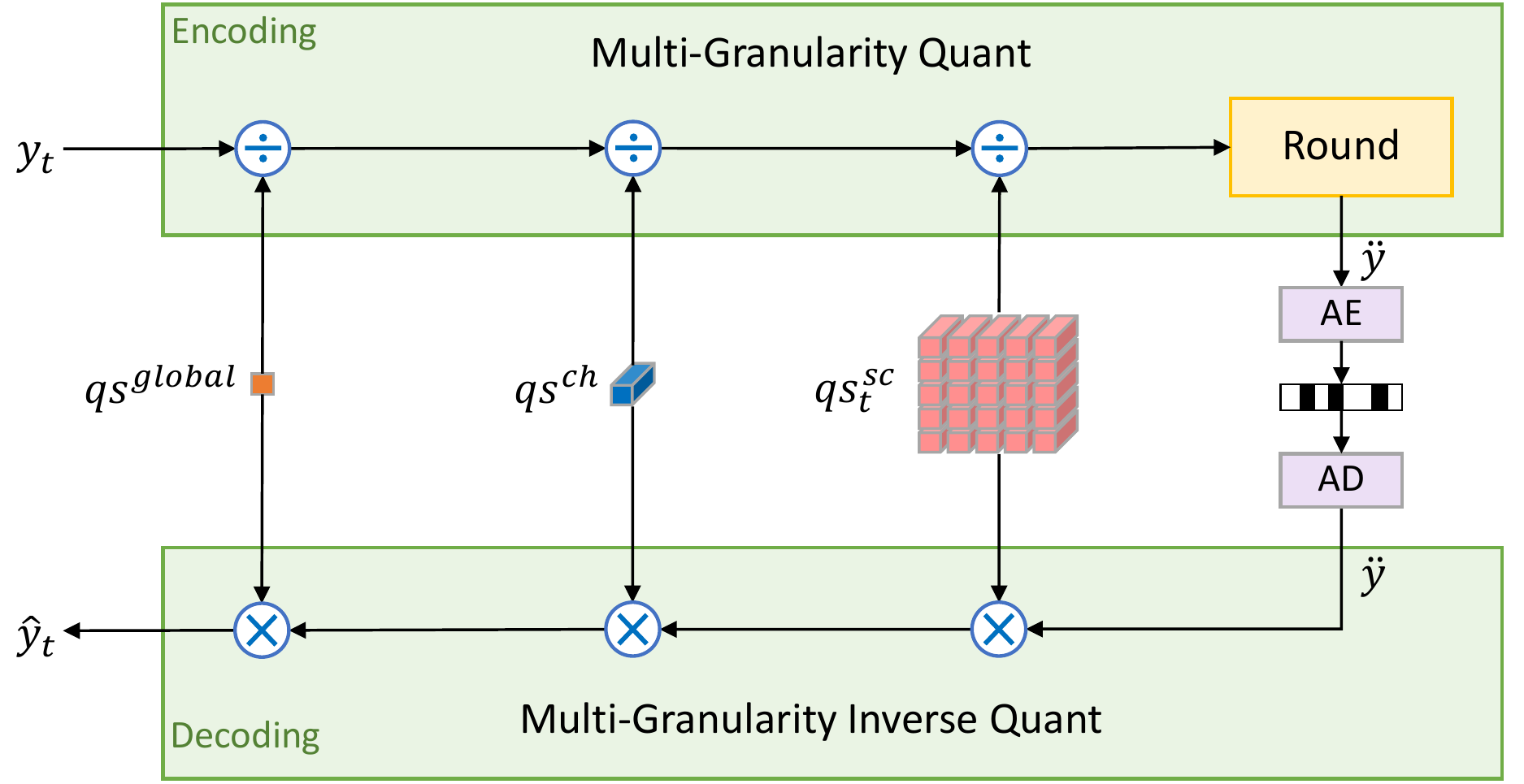}
  \vspace{-0.5cm}
  \caption{Multi-granularity quantization and the corresponding  inverse quantization.}
  \vspace{-0.5cm}
  \label{quantization}
\end{figure} 

As shown in Fig. \ref{fig_entropy_model}, the latent representation is split into two 
chunks along the channel dimension. During the first-step coding, the above branch in  Fig. \ref{fig_entropy_model} will code the even positions $\ddot{y}_{t, (i+j)\%2=0, k<\frac{C}{2}}$ in the first chunk. Simultaneously, the below branch codes the   odd positions $\ddot{y}_{t, (i+j)\%2=1, k\geq \frac{C}{2}}$ in the second chunk.
The uncoded positions (i.e., the white regions in Fig. \ref{fig_entropy_model}) in these two chunks are set as zero.
After the first-step coding, the coded $\ddot{y}_{t, (i+j)\%2=0, k<\frac{C}{2}}$ and $\ddot{y}_{t, (i+j)\%2=1, k\geq \frac{C}{2}}$ are fused together and then further generate the contexts for the second-step coding. During the second-step coding, the left positions in each chunk are coded. 
The above branch codes the odd positions $\ddot{y}_{t, (i+j)\%2=1, k<\frac{C}{2}}$ in the first chunk and the below branch codes the   even positions $\ddot{y}_{t, (i+j)\%2=0, k\geq \frac{C}{2}}$ in the second chunk. As shown in Fig.  \ref{fig_entropy_model}, this coding manner   enables that the second-step coding can benefit from the contexts from all positions.
When compared with original checkerboard  model \cite{he2021checkerboard}, the  scope of spatial context is doubled and results in more accurate prediction on distribution. It is noted that, for both steps, the first chunk and second chunk will be added and then sent to the arithmetic encoder. Thus,  our dual spatial prior will not bring any additional coding delay when compared with \cite{he2021checkerboard}. 

In addition, from the perspective of channel dimension, our dual spatial prior also mines the correlation across channels. The coded $\ddot{y}_{t, (i+j)\%2=0, k<\frac{C}{2}}$ during the first-step coding process can be also used as the condition for the coding of $\ddot{y}_{t, (i+j)\%2=0, k \geq \frac{C}{2}}$ during the second-step coding process. It is similar for the coding of $\ddot{y}_{t, (i+j)\%2=1, k\geq \frac{C}{2}}$  and $\ddot{y}_{t, (i+j)\%2=1, k < \frac{C}{2}}$, where the coding direction of channel is inverse. In a summary, our proposed dual spatial prior further squeezes the redundancy in $\ddot{y}_{t}$  by more efficiently exploiting the correlation across the spatial and channel positions.

\subsection{Rate Adjustment in Single Model}

It is a big pain that most existing neural codecs cannot handle  rate adjustment in single model. To achieve different rates, the model needs to be retrained by adjusting the weight in the RD loss. It will bring large training cost and model storage burden. Such shortcoming calls for a mechanism supporting neural codec to achieve wide rate range in single model. To this end, we propose an adaptive quantization mechanism   to enable this capability.

As shown in Fig. \ref{quantization}, our multi-granularity quantization involves three different kinds of quantization step (QP): the global QS $qs^{global}$, the channel-wise QS $qs^{ch}$, and the spatial-channel-wise QS $qs^{sc}$. The $qs^{global}$ is only a single  value and is set from the user input for controlling the target rate. As all positions take the same QS,  the $qs^{global}$  brings a coarse quantization effect. Thus, motivated by the channel attention mechanism \cite{hu2018squeeze}, we also design a modulator $qs^{ch}$ to  scale the QS at different channels because different channels carry information with different importance.  However, the different spatial positions also have different characteristics due to the various video contents. Thus, we follow \cite{Huang_2022_CVPR} and   learn the spatial-channel-wise  $qs^{sc}$ to achieve the precise adjustment on each  position.

\label{sub_sec_multirate}
\begin{figure}[t]
  \centering
  \includegraphics[width=1.05\linewidth]{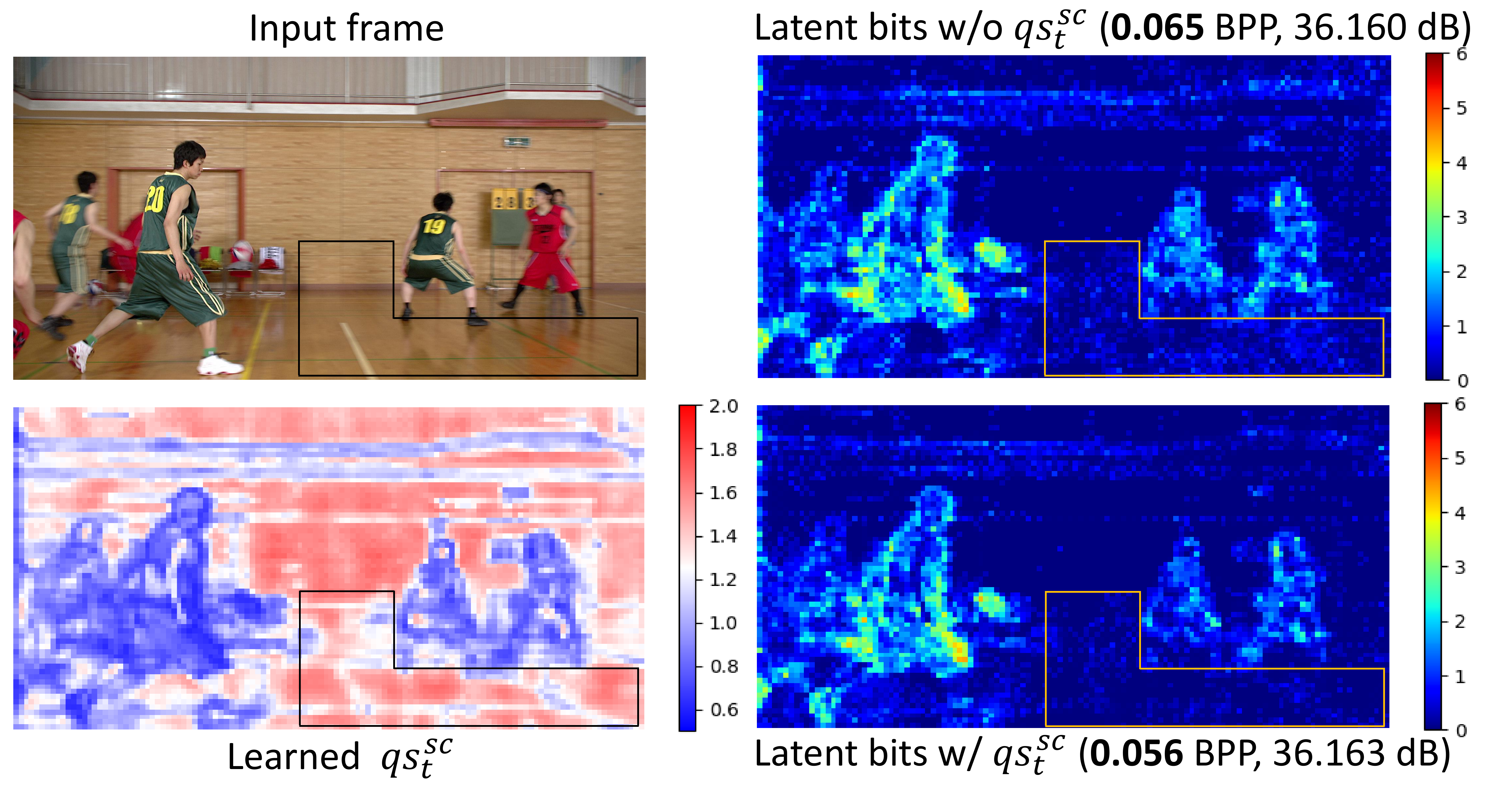}
  \vspace{-0.7cm}
  \caption{Effect of $qs^{sc}$. BPP means bits per pixel. In this example, $qs^{sc}$ brings 13.8\% BPP reduction under similar quality. Zoom in for better view.}
  \vspace{-0.5cm}
  \label{QStepVis}
\end{figure}

\begin{figure*}[t]
	\centering
	\includegraphics[width=1.02\linewidth]{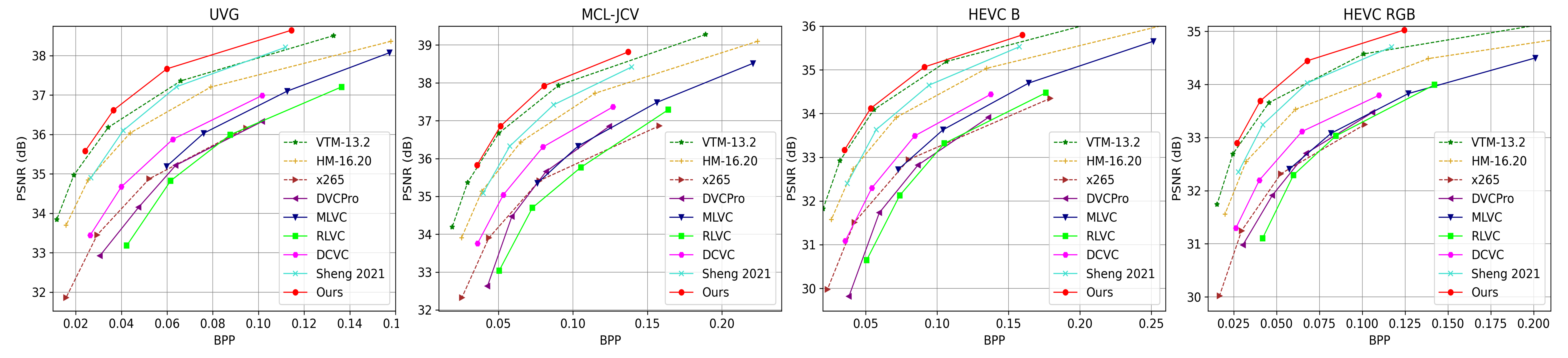} 
	\vspace{-0.8cm}
	\caption{PSNR and BPP curves.  
	}
	\label{rd_curve}
\end{figure*}

\begin{table*}[t]
  \centering
  \caption{BD-Rate (\%) comparison for PSNR. The anchor is VTM-13.2.}
   \renewcommand{\arraystretch}{1.25}
    \small
    \begin{tabular}{ccccccccc}
    \toprule[1.0pt]
                                         & UVG    & MCL-JCV  & HEVC B & HEVC C & HEVC D & HEVC E & HEVC RGB  & Average \\ \hline
    VTM-13.2                             & 0.0    & 0.0      & 0.0    & 0.0    & 0.0    & 0.0    & 0.0       & 0.0     \\ \hline
    HM-16.20                             & 40.5   & 45.4     & 40.4   & 40.9   & 36.0   & 46.2   & 42.1      & 41.6    \\ \hline
    x265                                 & 191.5  & 160.3    & 143.4  & 105.2  & 96.1   & 128.4  & 151.2     & 139.4   \\ \hline
    DVCPro \cite{lu2020end}              & 227.0  & 180.8    & 209.8  & 220.6  & 166.4  & 446.2  & 178.5     & 232.8   \\ \hline
    RLVC \cite{yang2021learning}         & 224.2  & 214.3    & 207.0  & 212.4  & 149.2  & 392.8  & 195.5     & 227.9   \\ \hline
    MLVC \cite{lin2020m}                 & 113.6  & 124.0    & 118.0  & 213.7  & 166.5  & 237.6  & 151.2     & 160.7   \\ \hline
    DCVC \cite{li2021deep}               & 126.1  & 98.2     & 115.0  & 150.8  & 109.6  & 266.2  & 109.6     & 139.4   \\ \hline
    Sheng 2021 \cite{sheng2021temporal}  & 17.1   & 30.6     & 28.5   & 60.5   & 27.8   & 67.3   & 17.9      & 35.7    \\ \hline
    Ours                                 & --18.2 & --6.4    & --5.1  & 15.0   & --8.9  & 7.1    & --16.4    & --4.7   \\
    \bottomrule[1.0pt]
    \end{tabular}%
  \label{tab_res_psnr}%
\end{table*}%

The $qs^{sc}$ is generated by our entropy model, as shown in Fig. \ref{fig_entropy_model}. It is noted that, for each  frame, it is dynamically changed to adapt the video contents. Such  design not only  helps us  achieve  smooth rate adjustment but also improves the final RD performance by content-adaptive bit allocation. The important information which is vital for the reconstruction or  is referenced by the coding of the subsequent frames will be allocated with  smaller QS. 
We show a visualization example of $qs^{sc}$ in Fig. \ref{QStepVis}. In this example, the model learns that the moving players are more important and produces smaller QS for these regions. By contrast, the backgrounds have larger QS, which brings considerable bitrate saving (as shown in region indicated by the yellow line).
The ablation study in section \ref{sub_sec_exp_ablation} shows  $qs^{sc}$ can achieve substantial improvements on multiple datasets.

During the decoding, the corresponding inverse quantization is applied. It is noted that, the $qs^{global}$ needs to be transmitted to the decoder along with the bit-stream. However, this overhead is negligible as only single  number is transmitted for each frame or video (flexible setting for user). The modulator $qs^{ch}$ belongs to a part of our neural codec and is learned during the training.


\section{Experimental Results}
\label{sec_experiments}

\subsection{Experimental Setup}
\label{sub_sec_exp_setup}

\textbf{\quad \  Datasets.} We use Vimeo-90k \cite{xue2019video} for training. The videos are randomly cropped into 256x256 patches. 
For testing, we use the same test videos as \cite{sheng2021temporal}. All these test videos are widely used in the evaluation of traditional and neural video codecs, including HEVC Class B, C, D, E, and RGB. In addition, the 1080p videos
from UVG \cite{mercat2020uvg} and MCL-JCV \cite{wang2016mcl} datasets are also tested.

\textbf{Test conditions.} We test 96 frames for each video. The intra period is set to 32 rather than 10 or 12. 
The reason is that intra period 32 gets closer to the practical usage in the real applications. For example, when compared with intra period 12, intra period 32 has an average of 23.8\% \cite{sheng2021temporal} bitrate saving for HM. 
We follow the low delay encoding settings as most existing works \cite{lu2019dvc,lu2020end, li2021deep,sheng2021temporal}. The compression ratio is measured by BD-Rate \cite{bjontegaard2001calculation}, where negative numbers indicate bitrate saving and positive numbers indicate bitrate increase.

Besides x265 \cite{x265} (\textit{veryslow} preset is used), our benchmarks include  HM-16.20 \cite{HM} and VTM-13.2 \cite{VTM}, which represent the best encoder of H.265 and H.266, respectively. 
For HM and VTM, we follow \cite{sheng2021temporal} and use the configuration with the highest compression ratio. 
We also compare  previous SOTA neural video codecs including DVCPro \cite{lu2020end}, MLVC \cite{lin2020m}, RLVC \cite{yang2021learning}, DCVC \cite{li2021deep}, as well as Sheng 2021 \cite{sheng2021temporal}. 

\begin{table*}[t]
    \centering
    \caption{BD-Rate  (\%) comparison for MS-SSIM. The anchor is VTM-13.2.}
    \renewcommand{\arraystretch}{1.25}
    \small
      \begin{tabular}{ccccccccc}
      \toprule[1.0pt]
                                          & UVG    & MCL-JCV & HEVC B & HEVC C & HEVC D & HEVC E & HEVC RGB  & Average \\ \hline
      VTM-13.2                            & 0.0    & 0.0     & 0.0    & 0.0    & 0.0    & 0.0    & 0.0       & 0.0     \\ \hline
      HM-16.20                            & 36.9   & 43.7    & 36.7   & 38.7   & 34.9   & 40.5   & 37.2      & 38.4    \\ \hline
      x265                                & 150.5  & 137.6   & 129.3  & 109.5  & 101.8  & 109.0  & 121.9     & 122.8   \\ \hline
      DVCPro \cite{lu2020end}             & 68.1   & 37.8    & 61.7   & 59.1   & 23.9   & 212.5  & 57.3      & 74.3    \\ \hline
      RLVC \cite{yang2021learning}        & 83.9   & 72.1    & 66.6   & 76.5   & 34.1   & 268.4  & 60.5      & 94.6    \\ \hline
      DCVC \cite{li2021deep}              & 33.6   & 4.7     & 31.0   & 22.8   & 1.2    & 124.7  & 36.5      & 36.4    \\ \hline
      Sheng 2021 \cite{sheng2021temporal} & --10.1 & --24.4  & --24.1 & --23.3 & --37.3 & --8.0  & --25.9    & --21.9  \\ \hline
      Ours                                & --35.1 & --46.8  & --48.1 & --44.6 & --55.7 & --47.5 & --47.0    & --46.4  \\
      \bottomrule[1.0pt]
      \end{tabular}%
    \label{tab_res_ssim}%
  \end{table*}%

\textbf{Implementation and training details.} The entropy model and quantization for the latent representation of motion vector follow   those of ${y}_t$. The only difference is the input of entropy model. In the coding of motion vector, the inputs are the corresponding hyper prior and the latent prior, i.e., the quantized latent representation of motion vector from the previous frame. There is no temporal context prior for the coding of motion vector because  the generation of temporal context depends on the decoded motion vector.  
In addition, as we target at single model handling multiple rates, we also train  a neural image codec supporting such capability for intra coding. 

During the training, the loss function includes the distortion and rate: $Loss=\lambda \cdot D + R$.
$D$ refers to the distortion between the input frame  and the reconstructed frame. The distortion can be  $L_2$ loss  or MS-SSIM~\cite{wang2003multiscale} for different visual targets. $R$ represents the bits used for encoding $\ddot{y}_t$ and the quantized latent representation of   motion vector, both associated with the bits used for encoding their corresponding  hyper prior.

We adopt the multi-stage training, same as \cite{sheng2021temporal}. In addition, to train single model supporting  rate adjustment, we use different $\lambda$ values in different optimization steps. For simplifying the training process, 4  $\lambda$ values (85, 170, 380, 840) are  used. During the training, 4  $qs^{global}$ values will be learned via the RD loss with each corresponding $\lambda$ value.
We trained image and video models separately and they have different $qs^{global}$ values.
It is noted that, although we only use 4 $\lambda$ values during the training, the model still can  achieve smooth rate adjustment by manually adjusting the $qs^{global}$ during the testing, and the corresponding study is presented  in Section \ref{sub_sec_exp_rc}.

\subsection{Comparisons with Previous SOTA Methods}
\label{sub_sec_exp_results}

Table \ref{tab_res_psnr} and Table \ref{tab_res_ssim} show the BD-Rate (\%) comparisons in terms  of PSNR and MS-SSIM, respectively. 
The best traditional codec VTM is used as anchor. 
From Table \ref{tab_res_psnr}, we can find that our neural codec achieves an average of 4.7\% bitrate saving over VTM on all datasets. By contrast, the second best method Sheng 2021 \cite{sheng2021temporal} is far behind than VTM  and has 35.7\% bitrate increase.  To the best of our knowledge, this is the first end-to-end  neural video codec that outperforms VTM using the highest compression ratio configuration, which is an important milestone in the development of neural video codec. In particular,  our neural codec performs better for 1080p videos (HEVC B, HEVC RGB, UVG, MCL-JCV). 
 Fig. \ref{rd_curve} shows  RD curves on these datasets. We can find our codec consumes the least bits under the same quality.
These results verify the effectiveness of our entropy model on exploiting the correlation among the volumed video data. In addition, the high-resolution video  will be more popular in the future and the advantage of our neural codec will be more obvious.
When oriented to MS-SSIM, our neural video codec  has larger improvement. As shown in Table \ref{tab_res_ssim}, we achieve an average of 46.4\% bitrate saving over VTM on all datasets. 

\subsection{Ablation Study}
\label{sub_sec_exp_ablation}

To verify the effectiveness of each component, we conduct a comprehensive ablation study. We analyze the effect of  entropy model input, our dual spatial prior, the  multi-granularity quantization, as well as the design on frame generator. For simplification, we only use HEVC testsets in ablation study. The comparisons are measured by BD-Rate (\%).

\begin{table}[t]
    \caption{Effect of entropy model inputs.  }
    \centering
    \scalebox{0.82}{
    \renewcommand{\arraystretch}{1.25}
        \begin{tabular}{ccccccccc}
            \toprule[1.0pt]
       Hyper        &    Temporal       &  Latent        &         \multirow{2}*{B}        & \multirow{2}*{C}         & \multirow{2}*{D}          & \multirow{2}*{E}          & \multirow{2}*{RGB}          & \multirow{2}*{Average}     \\
       prior        &     context prior        &  prior                &                                            &                                       &                       &              &               &                         \\ \hline
            \Checkmark   &\Checkmark   &\Checkmark    & 0.0     & 0.0     & 0.0      & 0.0      & 0.0        & 0.0     \\ \hline
            \Checkmark   &\Checkmark   &\XSolidBrush  & 9.5     & 5.6     & 6.4      & 18.4     & 10.7       & 10.1    \\ \hline
            \Checkmark   &\XSolidBrush &\Checkmark    & 9.2     & 8.8     & 9.1      & 15.0     & 9.1        & 10.2    \\ \hline
            \XSolidBrush &\Checkmark   &\Checkmark    & 19.5    & 9.1     & 11.9     & 27.2     & 20.9       & 17.7    \\ \hline
            \Checkmark   &\XSolidBrush &\XSolidBrush  & 12.4    & 14.7    & 17.9     & 23.4     & 9.4        & 15.6    \\ \hline
            \XSolidBrush &\Checkmark   &\XSolidBrush  & 33.8    & 29.3    & 31.6     & 54.3     & 36.2       & 37.0    \\ \hline
            \XSolidBrush &\XSolidBrush &\Checkmark    & 19.5    & 19.3    & 19.2     & 21.1     & 17.3       & 19.3    \\
            \bottomrule[1.0pt]
        \end{tabular}
    }
    \label{tab_prior_fusion}
\end{table}
\textbf{The inputs of entropy model.} As shown in Fig. \ref{fig_entropy_model}, our entropy model contains  three different kinds of inputs: hyper prior, temporal context prior, and latent prior. Table \ref{tab_prior_fusion} compares the effectiveness of these inputs. When removing our  latent prior, there is 10.1\% bitrate increase. If only enabling one prior input, the first and second most  important inputs are hyper prior (15.6\%) and our  latent prior (19.3\%). From these comparisons, we can find that the hyper prior is still the most  important. But enriching the entropy model input via our latent prior also brings
significant bitrate saving.

\textbf{Different spatial priors.} Table \ref{tab_spatial_prior} compares the effect of different spatial prior modules on exploring the inner correlation in $\ddot{y}_t$. The tested spatial context models include our proposed dual spatial prior, checkerboard  prior \cite{he2021checkerboard}, and  the parallel-unfriendly auto-regressive prior \cite{minnen2018joint}. From this table, we can find that  our dual spatial prior could bring 14.1\% bitrate saving. And there is  6.2\% improvement over the  checkerboard  prior. This verifies the benefit of enlarging the scope of spatial context and exploiting the cross-channel correlation. In addition, we also find that the auto-regressive  prior could further save the bitrate by a large margin (12.0\%). However, considering the very slow encoding and decoding speed, we still do not adopt it in our neural codec. 

\begin{table}[t]
\caption{Effect of different spatial priors. }
\centering
\scalebox{0.82}{
\renewcommand{\arraystretch}{1.25}
\begin{tabular}{ccccccc}
\toprule[1.0pt]
                             & B       & C       & D        & E        & RGB        & Average \\ \hline
Proposed dual spatial prior  & 0.0     & 0.0     & 0.0      & 0.0      & 0.0        & 0.0     \\ \hline
Checkerboard  prior          & 5.6     & 4.1     & 6.0      & 11.0     & 5.5        & 6.2     \\ \hline
No spatial prior             & 11.9    & 11.1    & 11.6     & 29.7     & 6.4        & 14.1    \\ \hline
Auto-regressive prior        & --14.8  & --11.4  & --11.2   & --5.6    & --16.9     & --12.0  \\
\bottomrule[1.0pt]
\end{tabular}
}
 \vspace{-0.3cm}
\label{tab_spatial_prior}
\end{table}

\textbf{ Multi-granularity quantization.} 
As introduced in Section \ref{sub_sec_multirate}, the multi-granularity quantization powered by our entropy model not only  helps us  achieve
rate adjustment in single model but also improves the final RD performance by content-adaptive bit allocation. 
Table \ref{tab_multi_rate} shows the BD-rate comparison. From this table, we can find that there is 11.8\% loss if disabling the whole multi-granularity quantization. 
When only removing the spatial-channel-wise quantization step $qs_t^{sc}$, there is a significant 8.9\% bit increase.  It shows that the $qs_t^{sc}$ play an important role in  multi-granularity quantization and it is necessary to design a dynamic bit allocation which can adapt to various video  contents.   

\begin{table}[t]
\caption{RD comparison of multi-granularity quantization. }
\centering
\scalebox{0.8}{
\renewcommand{\arraystretch}{1.4}
\begin{tabular}{ccccccc}
\toprule[1.0pt]
                               & B       & C       & D        & E        & RGB        & Average \\ \hline
Multi-granularity quantization     & 0.0     & 0.0     & 0.0      & 0.0      & 0.0        & 0.0     \\ \hline
w/o $qs_t^{sc}$            & 9.7     & 10.1     & 10.6      & 8.3      & 5.6        & 8.9     \\ \hline
w/o multi-granularity quantization      & 9.9     & 15.1    & 14.6     & 11.7     & 7.8        & 11.8    \\
\bottomrule[1.0pt]
\end{tabular}
}
\label{tab_multi_rate}
\end{table}

\textbf{Frame generator design.} To improve the  generation ability of neural codec,  our frame generator  uses W-Net based structure to enlarge the receptive field.  To verify its effectiveness, we  compare the frame generator using different networks, as shown in Table \ref{tab_recon_generation}. We compare W-Net, U-Net, and  residual block \cite{li2021deep,sheng2021temporal}  based structures. From this table, we can find that W-Net 
has more than 10\% bitrate saving than the plain network using  residual blocks. 
The W-Net which ties two U-Nets is also 7.9\% better than the single U-Net. Thus, 
it is possible that larger bitrate saving can be achieved by using more  U-Nets. However, considering the complexity, we currently use W-Net. 

\begin{table}[t]
\caption{Study on different frame generators. }
\centering
\scalebox{0.85}{
\renewcommand{\arraystretch}{1.2}
\begin{tabular}{ccccccc}
\toprule[1.0pt]
                  & B       & C       & D        & E        & RGB        & Average \\ \hline
W-Net             & 0.0     & 0.0     & 0.0      & 0.0      & 0.0        & 0.0     \\ \hline
U-Net             & 6.3     & 7.6     & 9.1      & 7.7      & 8.6        & 7.9     \\ \hline
2 Residual Blocks \cite{li2021deep,sheng2021temporal} & 11.7    & 9.9     & 11.0     & 13.4     & 9.1        & 11.0    \\ \hline
1 Residual Block  & 12.5    & 12.5    & 13.4     & 14.5     & 10.9       & 12.8    \\
\bottomrule[1.0pt]
\end{tabular}
}
\vspace{-0.3cm}
\label{tab_recon_generation}
\end{table}

\subsection{Smooth Rate Adjustment in Single Model}
\label{sub_sec_exp_rc}
\begin{figure}[t]
  \centering
  \includegraphics[width=\linewidth]{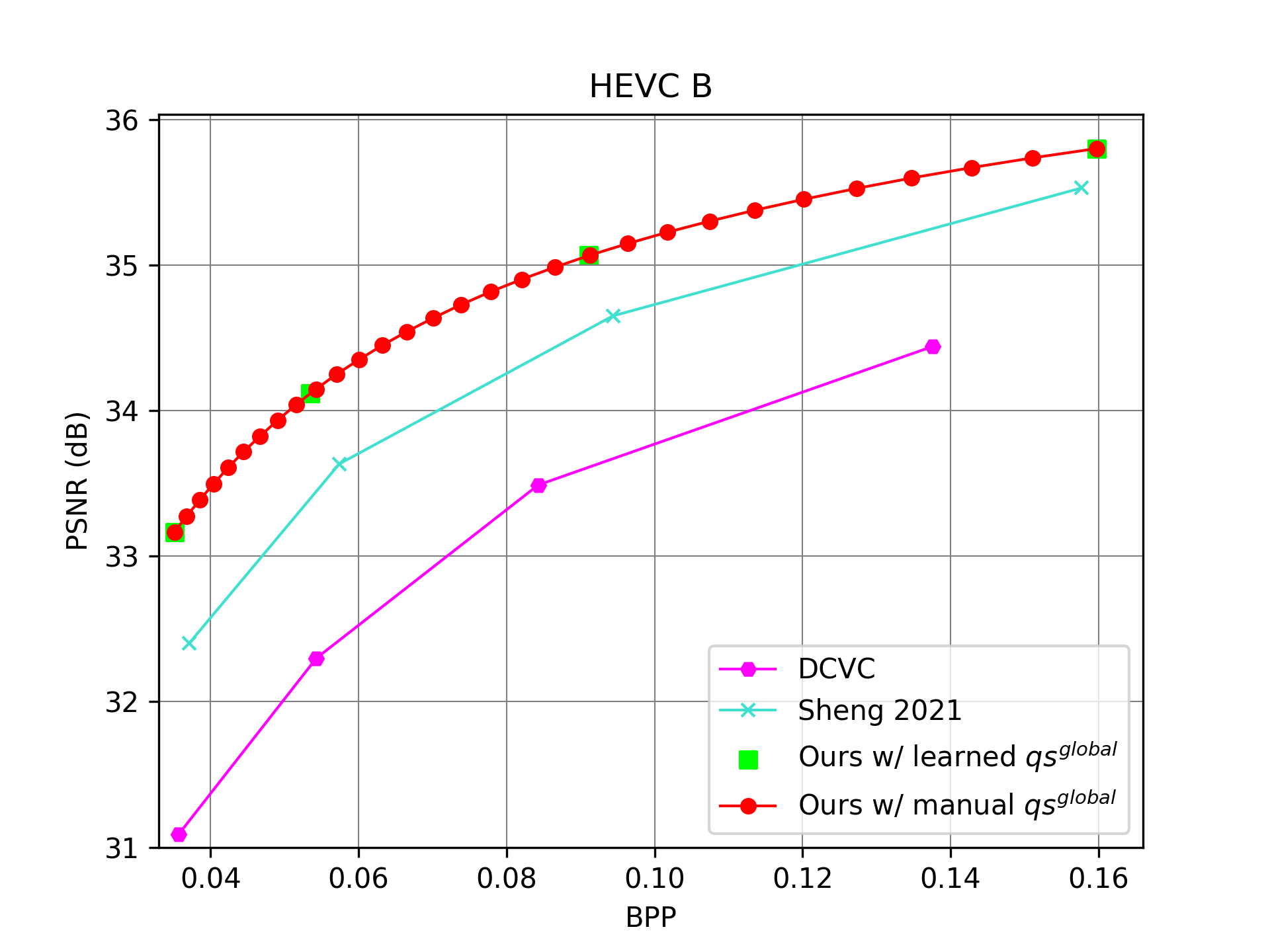}
  \vspace{-0.8cm}
  \caption{Rate adjustment in single model. DCVC \cite{li2021deep} and Sheng 2021 \cite{sheng2021temporal} need separate model for each rate point.}
  \label{fig_fine_rc}
\end{figure}

Table \ref{tab_multi_rate} shows the RD improvement of  multi-granularity quantization. This sub-section investigates  whether our multi-granularity quantization can achieve smooth rate adjustment. 
For our codec,  the global quantization step  $qs^{global}$ can be flexibly adjusted  during the testing.  It  serves  the similar role of quantization parameter  in traditional video codecs. 
Fig. \ref{fig_fine_rc}  shows the results of our codec using the learned 4 $qs^{global}$ values which are guided by RD loss using the 4 $\lambda$ values during the training. In addition, we also manually generate 30 $qs^{global}$ values by the interpolation between the maximum and minimum of the learned $qs^{global}$ values.  
From the results at the 30 rate points, we can find that  our single model  could achieve smooth rate adjustment without any outlier. 
 By contrast,
DCVC \cite{li2021deep} and Sheng 2021 \cite{sheng2021temporal} need  different models for each rate point. 
 

\subsection{Model Complexity}
\label{sub_sec_complexity}
We compare the model complexity in model size, MACs (multiply–accumulate operations),   encoding time, and decoding time with previous SOTA neural video codecs, as shown in Table \ref{tab_complexity}. We use 1080p frame as input to measure these numbers. For the encoding/decoding time, we measure the time on NVIDIA V100 GPU, including the time of writing to and reading from bitstream as in \cite{sheng2021temporal}. As aforementioned, our model supports rate adjustment in single model. Thus,
we significantly reduce the model training and storage burden.
Because DCVC \cite{li2021deep} uses the parallel-unfriendly auto-regression prior model,  its encoding/decoding time is quite slow. By
contrast, \cite{sheng2021temporal} and our codec are much faster.
Compare with \cite{sheng2021temporal}, our encoding/decoding time is increased a little. However,  the compression ratio is pushed into a new height (from Sheng 2021 surpassing HM  to ours surpassing VTM). We believe this is a price worth paying. 

\begin{table}[t]
\caption{Complexity comparison.}
\centering
\scalebox{0.85}{
\renewcommand{\arraystretch}{1.25}
\begin{tabular}{cccccc}
\toprule[1.0pt]
                                    & Model  size  &  MACs        & Encoding time & Decoding     time        \\ \hline
DCVC \cite{li2021deep}              & 35.2MB $\times N$ & 2.4T  & 12,260ms      & 35,590ms            \\ \hline
Sheng 2021 \cite{sheng2021temporal} & 40.9MB $\times N$ & 2.9T   & 880ms         & 470ms        \\ \hline
Ours                                & 67.0MB $\times 1$ & 3.3T   & 986ms         & 525ms    \\
\bottomrule[1.0pt]
\end{tabular}
}
\\
\raggedright\footnotesize{ \, Note: $N$ is the number of rate points.  1080p is used for test.}
   \vspace{-0.3cm}
\label{tab_complexity}
\end{table}

\section{Conclusion}
\label{sec_conclusions}
In this paper, we have presented how to design an efficient entropy model which helps our neural codec not only  achieve the higher compression ratio than VTM but also smoothly adjust the rates in single model. 
In particular, the latent prior is proposed to enrich the input of entropy model. Via building the  the propagation chain, the model can exploit the correlation across the latent representation of multiple frames. 
The proposed dual spatial prior not only doubles the scope of spatial context in parallel-friendly manner but also further squeezes the redundancy across the channel dimension. In addition, our entropy model also generates the spatial-channel-wise quantization step. Such content-adaptive quantization mechanism helps  the codec cope with various video consents well. As the core element in the multi-granularity quantization, it not only helps achieve the smooth rate adjustment but also improves the final RD performance by dynamic bit allocation.  When compared with the best traditional codec VTM using the highest compression ratio configuration, an average of 18.2\% bitrate saving is achieved on 
UVG dataset, which  is an important milestone.

\newpage


\clearpage
\begin{appendices}

Appendices provide the supplementary material to our proposed hybrid spatial-temporal entropy  modelling for neural video compression. 

\section{Network Architecture}
\begin{figure} 
	\centering
	\includegraphics[width=1.0\linewidth]{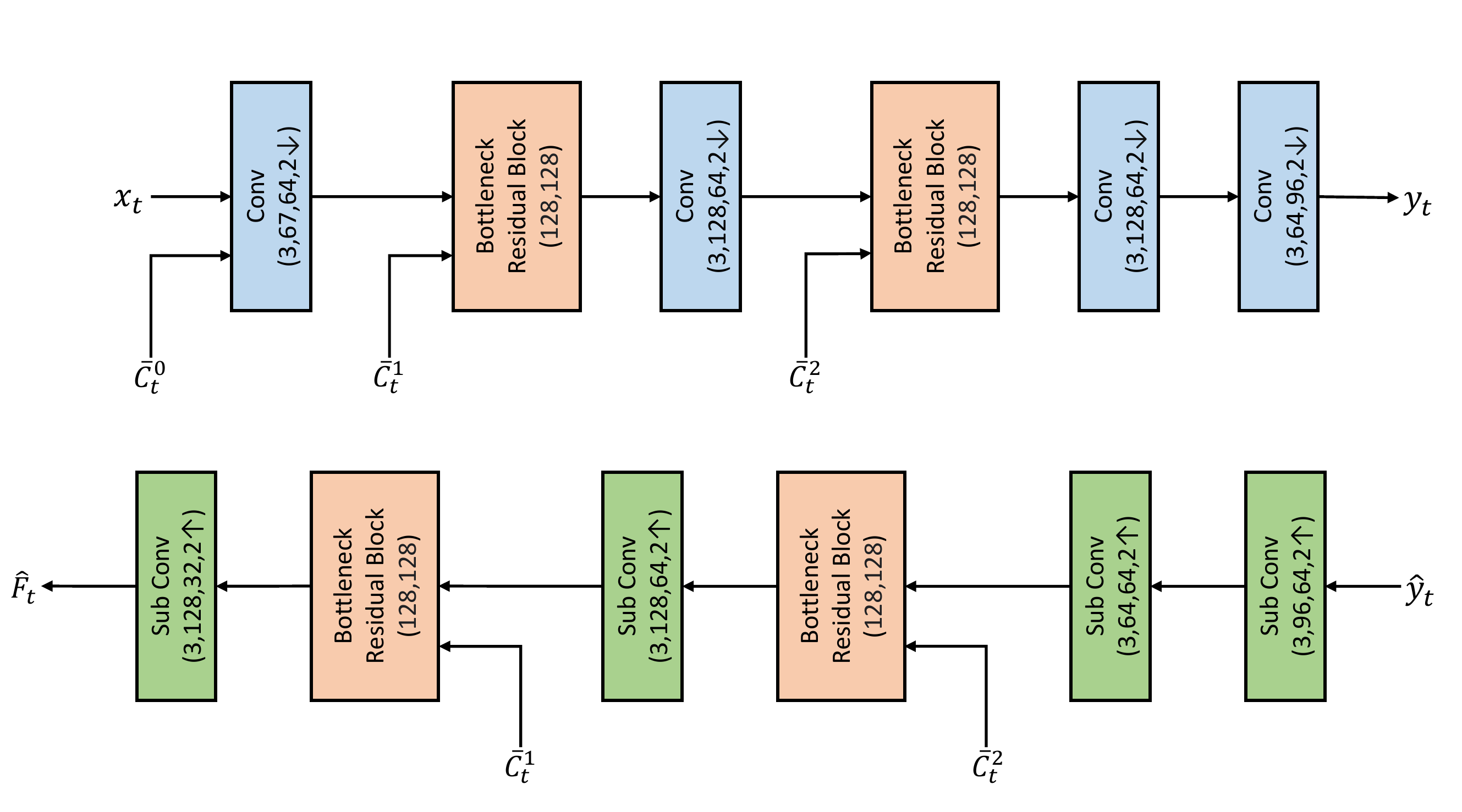}
	\caption{Structure of contextual encoder and decoder.  }
	\label{supp_contextual_encdec}
\end{figure}

\textbf{Contextual encoder and decoder.} The network design of our contextual encoder and decoder is shown in Fig. \ref{supp_contextual_encdec}. For  encoder, the  inputs are the current original frame $x_t$  and the multi-scale contexts {$\bar{C}^{0}_{t}$, $\bar{C}^{1}_{t}$,$\bar{C}^{2}_{t}$} at different  resolutions  (original, 2x downsampled, 4x downsampled) with channel 64. The output is the latent representation  $y_t$ with channel 96 at 16x downsampled resolution.
For decoder, the inputs contain the decoded latent representation  $\hat{y}_t$.  Conditioned on $\bar{C}^{1}_{t}$ and $\bar{C}^{2}_{t}$, the high-resolution feature $\hat{F}_t$  with channel 32 is decoded.  For the convolutional and sub-pixel convolutional layers in Fig. \ref{supp_contextual_encdec}, the $(K,Cin,Cout,S)$ indicate the kernel size, input channel number, output channel number, and stride, respectively. In Fig. \ref{supp_contextual_encdec}, the  bottleneck residual block  comes from \cite{sheng2021temporal}. The kernel size and stride of convolutional layer therein are 3 and 1, respectively. Thus,  we only show the input and output channel numbers for bottleneck residual block.

\begin{figure*} 
	\centering
	\includegraphics[width=0.9\linewidth]{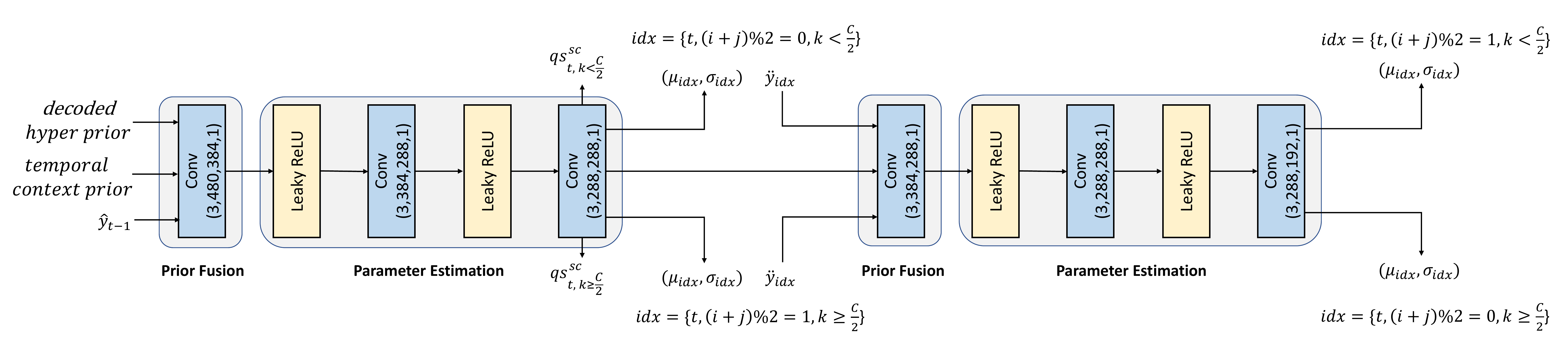} 
	\caption{ Structure of entropy model. }
	\label{supp_entropy_network}
\end{figure*}

\begin{figure} 
	\centering
	\includegraphics[width=0.7\linewidth]{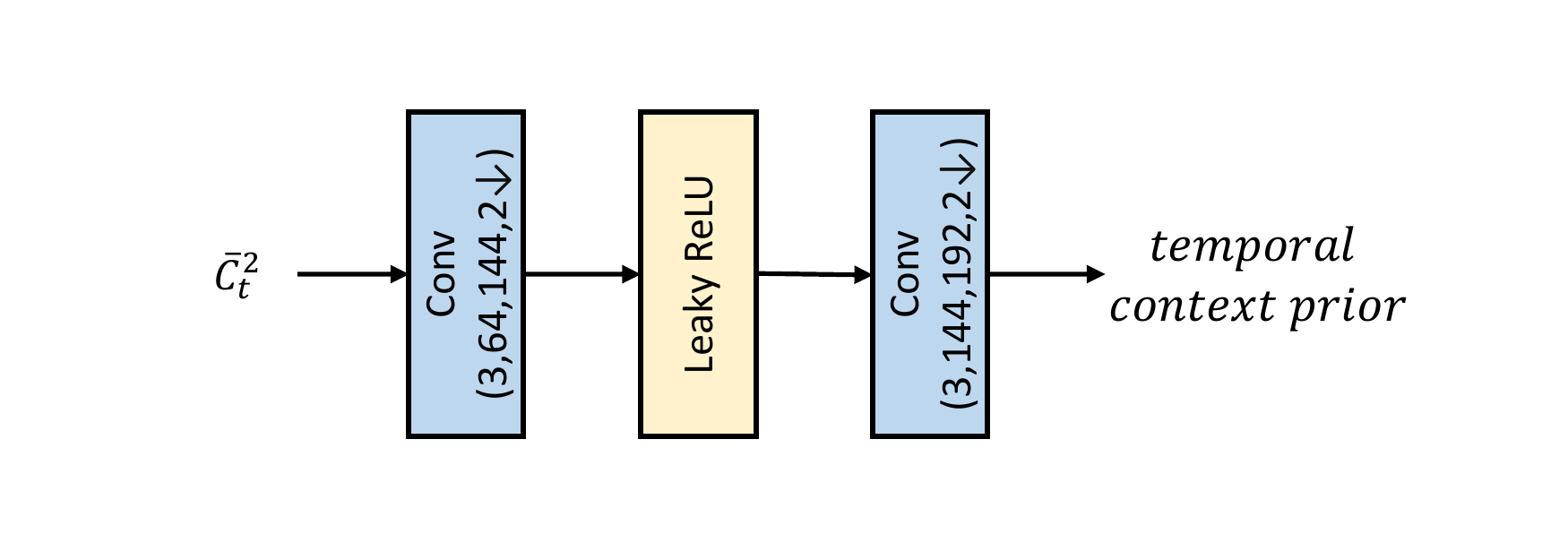}
	\caption{Structure of temporal context encoder.  }
	\label{supp_temporalprior}
\end{figure}

\begin{figure} 
	\centering
	\includegraphics[width=1.0\linewidth]{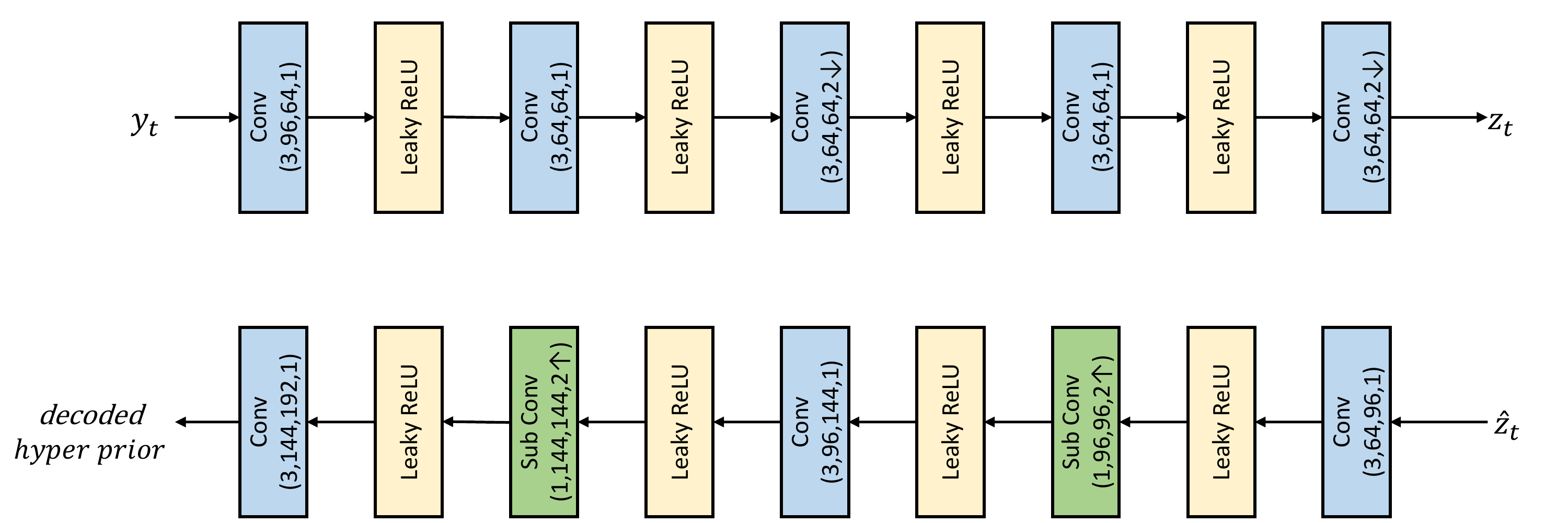}
	\caption{Structure of hyper prior encoder and decoder.  }
	\label{supp_hyper_encdec}
\end{figure}

\textbf{Entropy model.} Fig. \ref{supp_entropy_network} shows the detailed network structure of our entropy model.
The inputs include the the hyper prior with channel 192, the temporal context prior with channel 192, the latent prior (i.e., the decoded latent representation from the previous frame) with channel 96. The temporal context prior is generated by the temporal context encoder, as shown in Fig. \ref{supp_temporalprior}. The network structures of hyper prior encoder and decoder are shown in Fig. \ref{supp_hyper_encdec}.

In the first-step coding, the entropy model not only estimates the mean and scale values of the probability distribution for the quantized latent representations $\ddot{y}_{t, (i+j)\%2=0, k<\frac{C}{2}}$ and $\ddot{y}_{t, (i+j)\%2=1, k\geq \frac{C}{2}}$, but also generates the quantization step at spatial-channel-wise. During the second-step coding, the coded latent representations in previous step are  fused to the input, and then the entropy model estimates the the mean and scale values of the probability distribution for  $\ddot{y}_{t, (i+j)\%2=1, k<\frac{C}{2}}$ and  $\ddot{y}_{t, (i+j)\%2=0, k\geq \frac{C}{2}}$.

\begin{figure} 
	\centering
	\includegraphics[width=1\linewidth]{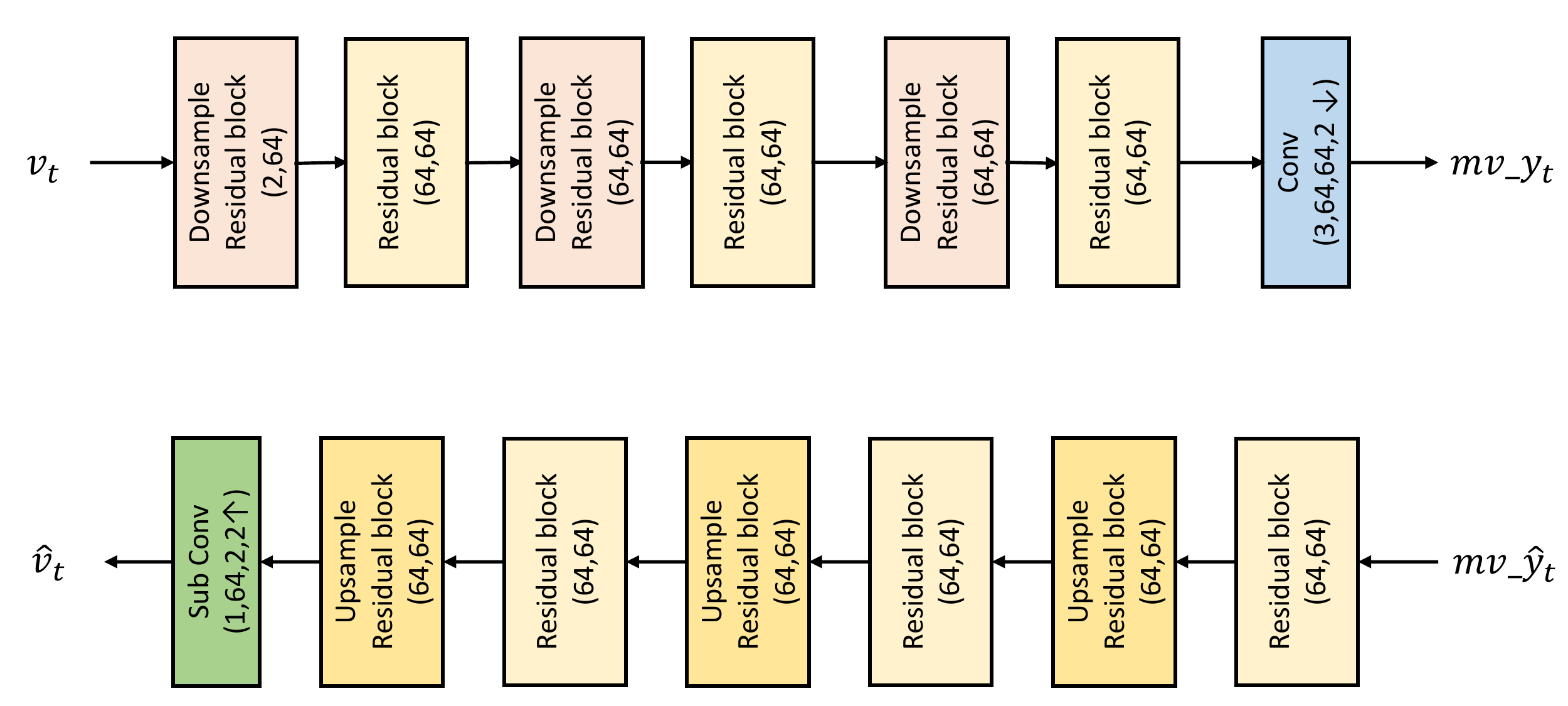}
	\caption{Structure of motion vector encoder and decoder.   }
	\label{supp_mv_encdec}
\end{figure}

\textbf{Motion vector encoder and decoder.} The encoder and decoder for motion vector are illustrated in  Fig. \ref{supp_mv_encdec}. 
For encoder, the input is the original motion vector with channel 2. The output is the latent representation at 16x downsampled resolution with channel 64. The decoder follows the inverse structure.
The multi-granularity quantization and entropy model for  ${mv\_y}_t$ follow   those of ${y}_t$. The only difference is the entropy model input. For motion vector, the inputs of entropy model are the corresponding hyper prior and the latent prior. In Fig. \ref{supp_mv_encdec}, the downsample and upsample residual blocks are from \cite{sheng2021temporal}.

\begin{figure} 
	\centering
	\includegraphics[width=0.85\linewidth]{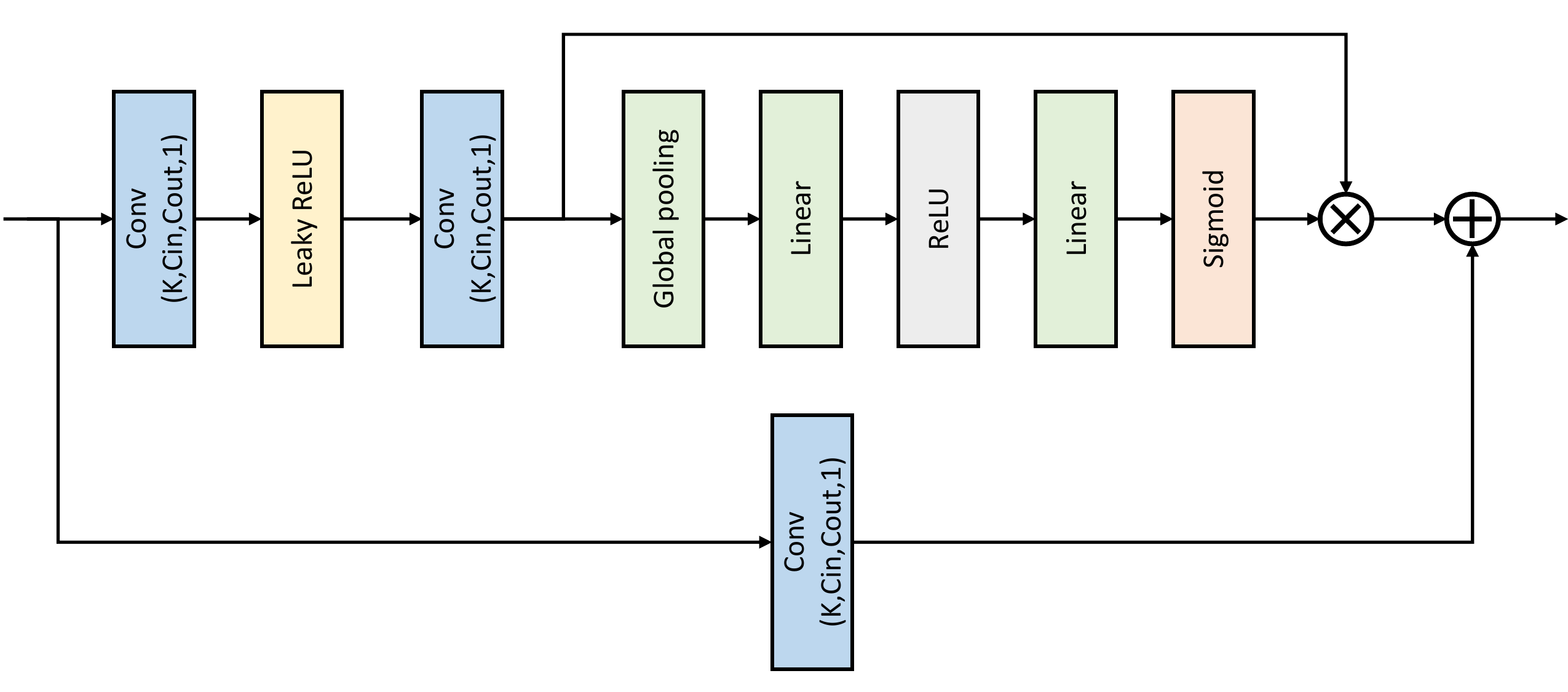}
	\caption{Structure of residual block with attention.  }
	\label{supp_resi_block_atten}
\end{figure}

\begin{figure} 
	\centering
	\includegraphics[width=1\linewidth]{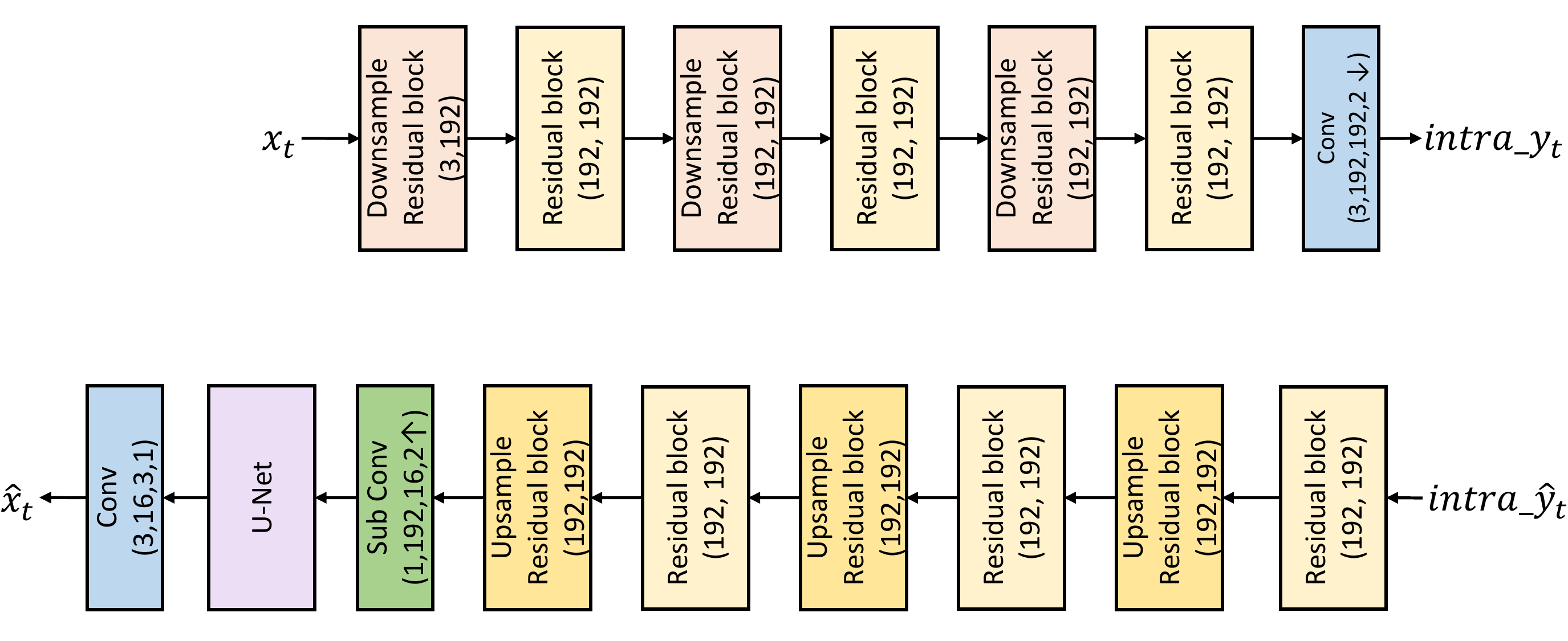}
	\caption{Structure of neural image codec.  }
	\label{supp_image}
\end{figure}

\begin{figure*} 
	\centering
	\includegraphics[width=0.9\linewidth]{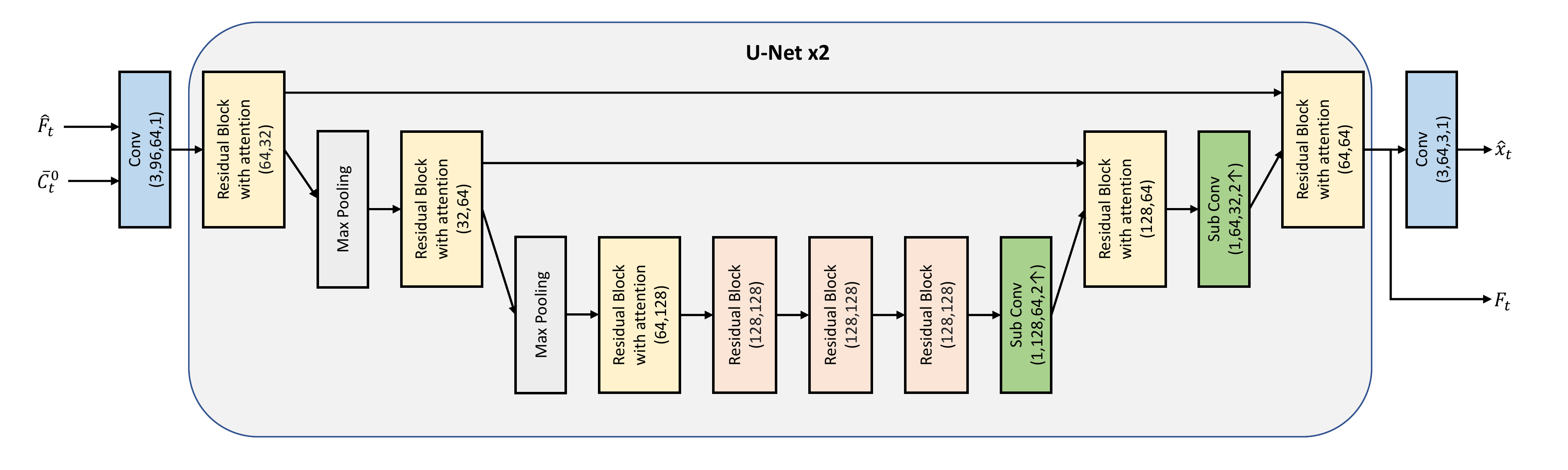} 
	\vspace{-0.4cm}
	\caption{ Structure of frame generator. }
	\label{supp_frame_generator}
\end{figure*}

\textbf{Frame generator.} Different from DCVC \cite{li2021deep} and \cite{sheng2021temporal} only using plain residual blocks, we proposing using the W-Net \cite{xia2017w} based  structure to build our frame generator. 
The W-Net can effectively enlarge the receptive field of model to improve the   generation ability of model. The detailed network structure of our frame generator is presented in Fig. \ref{supp_frame_generator}.
The inputs include the the high-resolution feature $\hat{F}_t$    with channel 32 and the context $\bar{C}^{0}_{t}$ at original resolution with channel 64.  The outputs are the final reconstructed frame  and the feature used by the next frame. In Fig. \ref{supp_frame_generator}, there are two same U-Nets to build the W-Net. The residual block with attention in the U-Net is illustrated in Fig.\ref{supp_resi_block_atten}.  

\textbf{Neural image codec.}  As we target at single model handling multiple rates, we also need  a neural image codec supporting such capability for intra coding. The structure of our neural image codec is presented in Fig. \ref{supp_image}. The corresponding multi-granularity quantization and the entropy model are similar with those of   ${y}_t$. The only difference is the input of entropy model. For neural image codec, the input of entropy model only includes the hyper prior. In Fig. \ref{supp_image}, we also use one U-Net to improve the generation ability of neural image codec and its network structure is similar to that in Fig. \ref{supp_frame_generator}. 
The compression ratio of our image codec is on par with that of Cheng 2020 \cite{cheng2020learned}, as shown in Table \ref{image_results}.

\section{Settings of Traditional Codecs}
For    x265 \cite{x265}, we use the \textit{veryslow} preset. We also compare
HM-16.20 \cite{HM} and VTM-13.2 \cite{VTM}, which represent the best encoder of H.265 and H.266, respectively. 
For HM and VTM, the low delay configuration with the highest compression ratio is used. 4 reference frames and 10-bit internal bit depth are used (settings in \textit{encoder\_lowdelay\_main\_rext.cfg} for HM and \textit{encoder\_lowdelay\_vtm.cfg } for VTM). The comparison in \cite{sheng2021temporal} has shown that, although the final distortion is measured in RGB domain, using YUV 444  as the internal color space  could boost the compression ratio. Thus, we also use this setting. More studies on codec setting can be found in \cite{sheng2021temporal}. The detailed settings of x265, HM, and VTM are:

\begin{itemize}
	\item {\bf x265}\par ffmpeg \par
	-pix\_fmt yuv420p \par
	-s \{{\em width}\}x\{{\em height}\}\par
	-framerate \{{\em frame rate}\}\par
	-i \{{\em input file name}\}\par
	-vframes \{{\em frame number}\}\par
	-c:v libx265\par
	-preset veryslow\par
	-tune zerolatency\par
	-x265-params \par
	``qp=\{{\em qp}\}:keyint=32:csv-log-level=1:\par
	csv=\{{\em csv\_path}\}:verbose=1:psnr=1"\par
	\{{\em output video file name}\} \par
\end{itemize}

\begin{itemize}
	\item {\bf HM} \par TAppEncoder \par
	-c encoder\_lowdelay\_main\_rext.cfg \par
	-\/-InputFile=\{{\em input file name}\}\par
	-\/-InputBitDepth=8 \par
	-\/-OutputBitDepth=8 \par
	-\/-OutputBitDepthC=8 \par
	-\/-InputChromaFormat=444\par
	-\/-FrameRate=\{{\em frame rate}\}\par
	-\/-DecodingRefreshType=2\par
	-\/-FramesToBeEncoded=\{{\em frame number}\}\par
	-\/-SourceWidth=\{{\em width}\}\par
	-\/-SourceHeight=\{{\em height}\}\par
	-\/-IntraPeriod=32\par
	-\/-QP=\{{\em qp}\}\par
	-\/-Level=6.2\par
	-\/-BitstreamFile=\{{\em bitstream file name}\}
\end{itemize}

\begin{itemize}
	\item {\bf VTM}\par EncoderApp\par
	-c encoder\_lowdelay\_vtm.cfg \par
	-\/-InputFile=\{{\em input file name}\}\par
	-\/-BitstreamFile=\{{\em bitstream file name}\}\par
	-\/-DecodingRefreshType=2\par
	-\/-InputBitDepth=8\par
	-\/-OutputBitDepth=8 \par
	-\/-OutputBitDepthC=8 \par
	-\/-InputChromaFormat=444\par
	-\/-FrameRate=\{{\em frame rate}\}\par
	-\/-FramesToBeEncoded=\{{\em frame number}\}\par
	-\/-SourceWidth=\{{\em width}\}\par
	-\/-SourceHeight=\{{\em height}\}\par
	-\/-IntraPeriod=32\par
	-\/-QP=\{{\em qp}\}\par
	-\/-Level=6.2\par
\end{itemize}

\begin{table*}[t]
	\centering
	\caption{Image codec BD-Rate (\%) comparison for PSNR.}
	\renewcommand{\arraystretch}{1.25}
	\small
	\begin{tabular}{ccccccccc}
		\toprule[1.0pt]
		& UVG    & MCL-JCV    & HEVC B     & HEVC C  & HEVC D  & HEVC E   & HEVC RGB  & Average \\ \hline
		Cheng 2020 image codec \cite{cheng2020learned}          &  0.0    &    0.0     &     0.0    &   0.0   &   0.0   &   0.0    &  0.0    &   0.0     \\ \hline
		Our image codec                  &  -0.9    &   0.2    &     -1.9   &   -0.2  &   2.5  &   1.9  &  -1.2   &   0.1    \\
		\bottomrule[1.0pt]
	\end{tabular}
	\label{image_results}%
\end{table*}%

\section{Rate-Distortion Curves}
\begin{figure*}[t]
	\centering
	\includegraphics[width=0.75\linewidth]{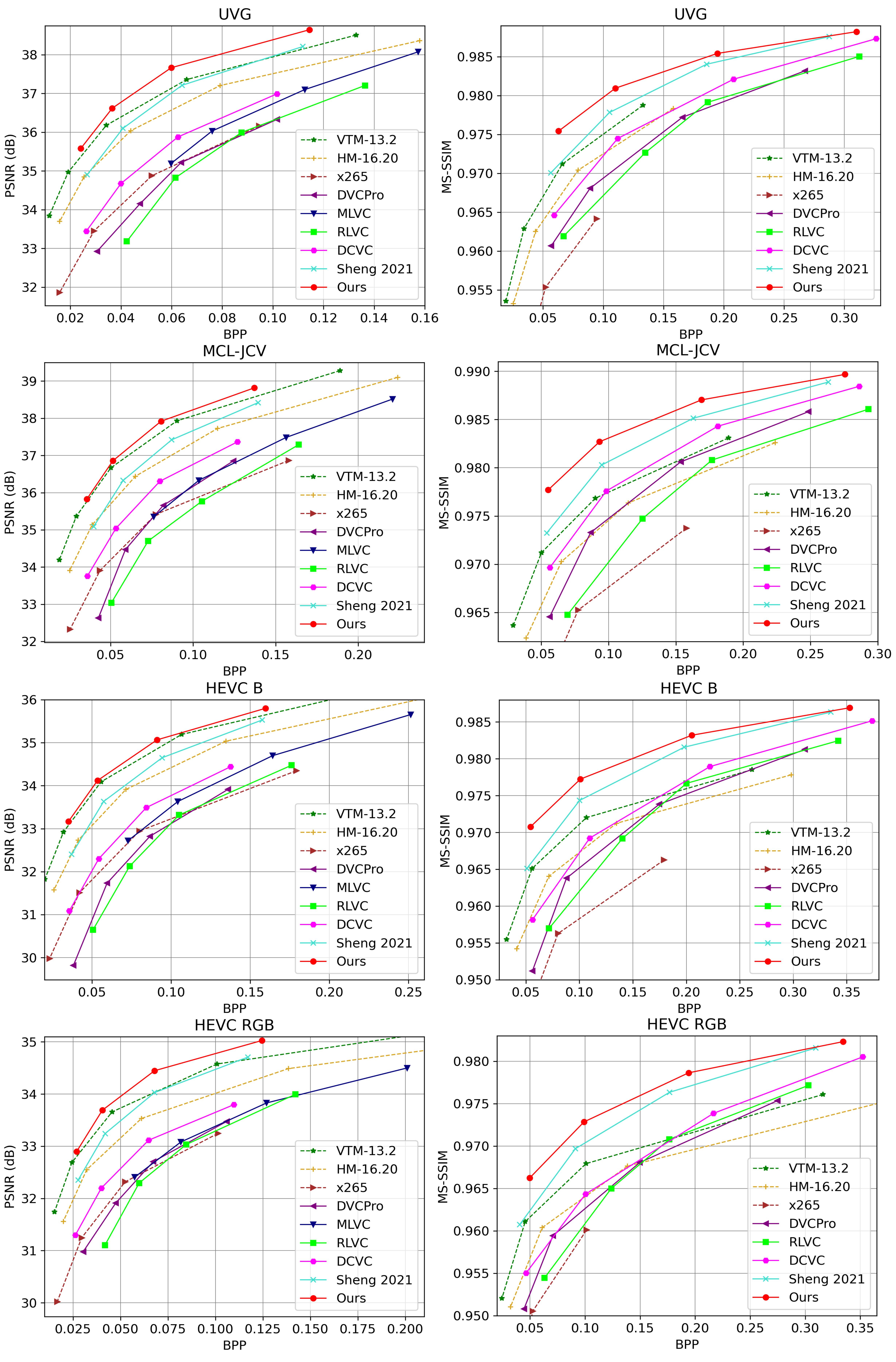}
	\vspace{-0.4cm}
	\caption{RD curves of UVG, MCL-JCV, HEVC B and  RGB.  }
	\label{supp_rd_curve}
\end{figure*}

\begin{figure*}[t]
	\centering
	\includegraphics[width=0.75\linewidth]{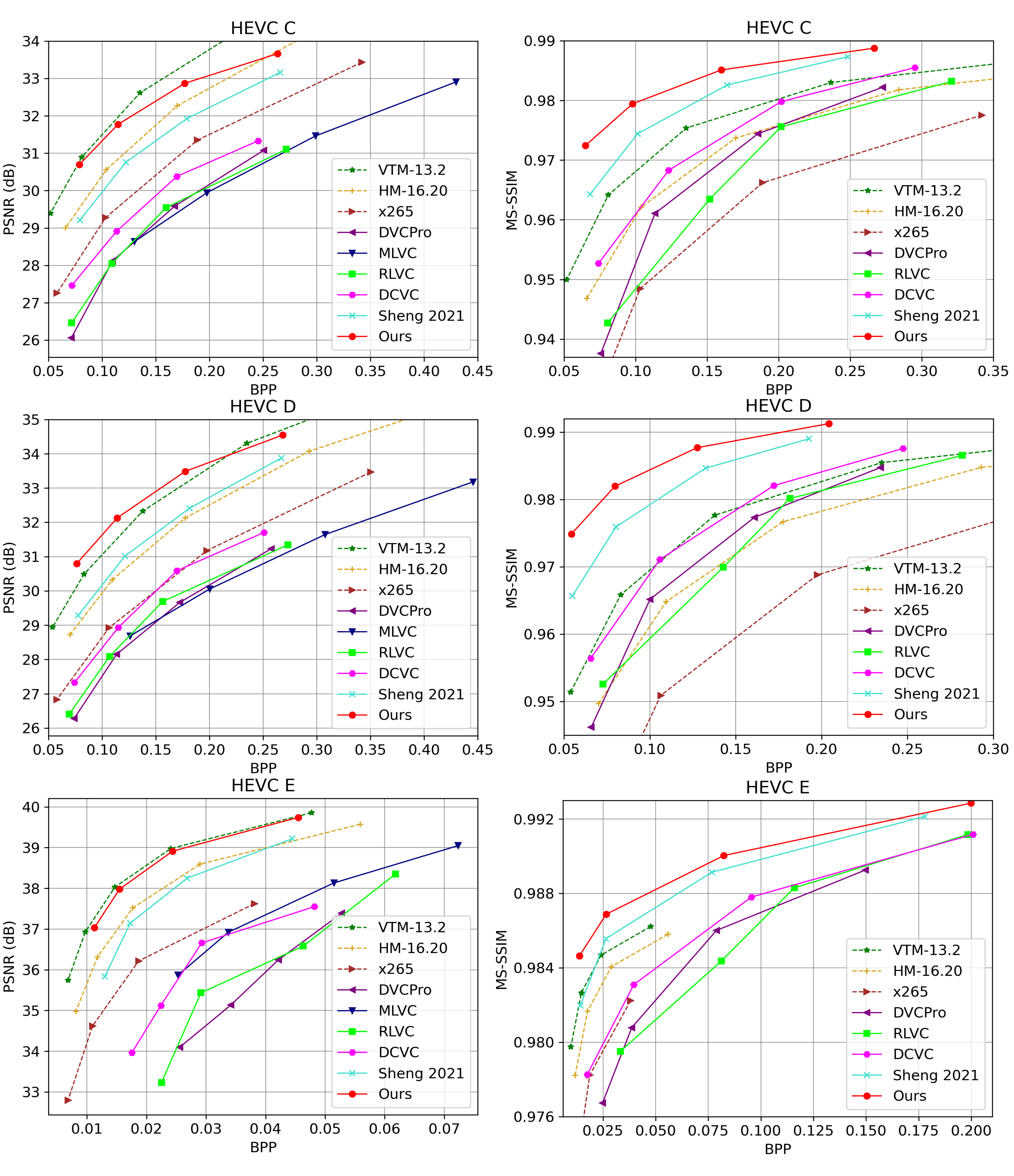}
	\vspace{-0.4cm}
	\caption{RD curves of HEVC C, D  and E.  }
	\label{supp_rd_curve2}
\end{figure*}

Fig. \ref{supp_rd_curve} and Fig. \ref{supp_rd_curve2} show the RD (rate-distortion) curves on each dataset, including both the PSNR and MS-SSIM results. From these results, we can find that our codec can achieve SOTA compression ratio in wide rate range.

\section{Comparison under Different Intra Period Settings}
Previous work Sheng 2021 \cite{sheng2021temporal} shows that intra period 12 is harmful to the compression ratio and is seldom used in the real applications. For example, when compared with intra period 12, intra period 32 has an average of 23.8\% bitrate saving for HM (Table 3 in the supplementary material of \cite{sheng2021temporal}). Thus, to get closer to the practical scenario, we follow [39] and also use intra period 32 in our paper.

But it is also important to evaluate our codec under intra period 12. Table \ref{tab_res_psnr_ip12} shows the corresponding BD-rate (\%) comparison in terms of PSNR. It is noted that, since the GOP size of VTM with *encoder\_lowdelay\_vtm* configuration is 8, it does not support intra period 12. Therefore, we configure VTM-13.2* by following \cite{sheng2021temporal} and use it as the anchor (more configuration details of VTM-13.2* can be found in \cite{sheng2021temporal}).
We can see that, under intra period 12, our codec also significantly outperforms all previous SOTA neural and traditional codecs. 

\begin{table*}[t]
	\centering
	\caption{BD-Rate (\%) comparison for PSNR under intra period 12.}
	\renewcommand{\arraystretch}{1.25}
	\small
	\begin{tabular}{ccccccccc}
		\toprule[1.0pt]
		& UVG    & MCL-JCV    & HEVC B     & HEVC C  & HEVC D  & HEVC E   & HEVC RGB  & Average \\ \hline
		VTM-13.2*                             &  0.0    &    0.0     &     0.0    &   0.0   &   0.0   &   0.0    &  0.0    &   0.0     \\ \hline
		HM-16.20                             &  8.3    &    15.3    &     20.1   &   13.3  &   11.6  &   22.4   &  13.2   &   14.9    \\ \hline
		x265                                 &  97.4   &    99.2    &     89.1   &   49.5  &   44.2  &   79.5   &  94.5   &   79.1    \\ \hline
		DVCPro \cite{lu2020end}              &  54.8   &    63.0    &     67.3   &   70.5  &   47.2  &   124.2  &  50.8   &   68.3    \\ \hline
		RLVC \cite{yang2021learning}         &  74.0   &    103.8   &     87.2   &   86.1  &   53.0  &   98.7   &  66.2   &   81.3    \\ \hline
		MLVC \cite{lin2020m}                 &  39.9   &    65.0    &     57.2   &   99.5  &   75.8  &   84.1   &  64.3   &   69.4    \\ \hline
		DCVC \cite{li2021deep}               &  21.0   &    28.1    &     32.5   &   44.8  &   25.2  &   66.8   &  22.9   &   34.5    \\ \hline
		Sheng 2021 \cite{sheng2021temporal}  &  -20.4  &    -1.6    &     -1.0   &   15.7  &   -3.4  &   7.8    &  -13.7  &   -2.4    \\ \hline
		Ours                                 &  -38.4  &    -24.2   &     -19.9  &   -10.5 &   -23.9 &   -18.7  &  -34.2  &   -24.3   \\
		\bottomrule[1.0pt]
	\end{tabular}
	\label{tab_res_psnr_ip12}%
\end{table*}%

\section{Visual Comparisons}
\begin{figure*}[t]
	\centering
	\includegraphics[width=1\linewidth]{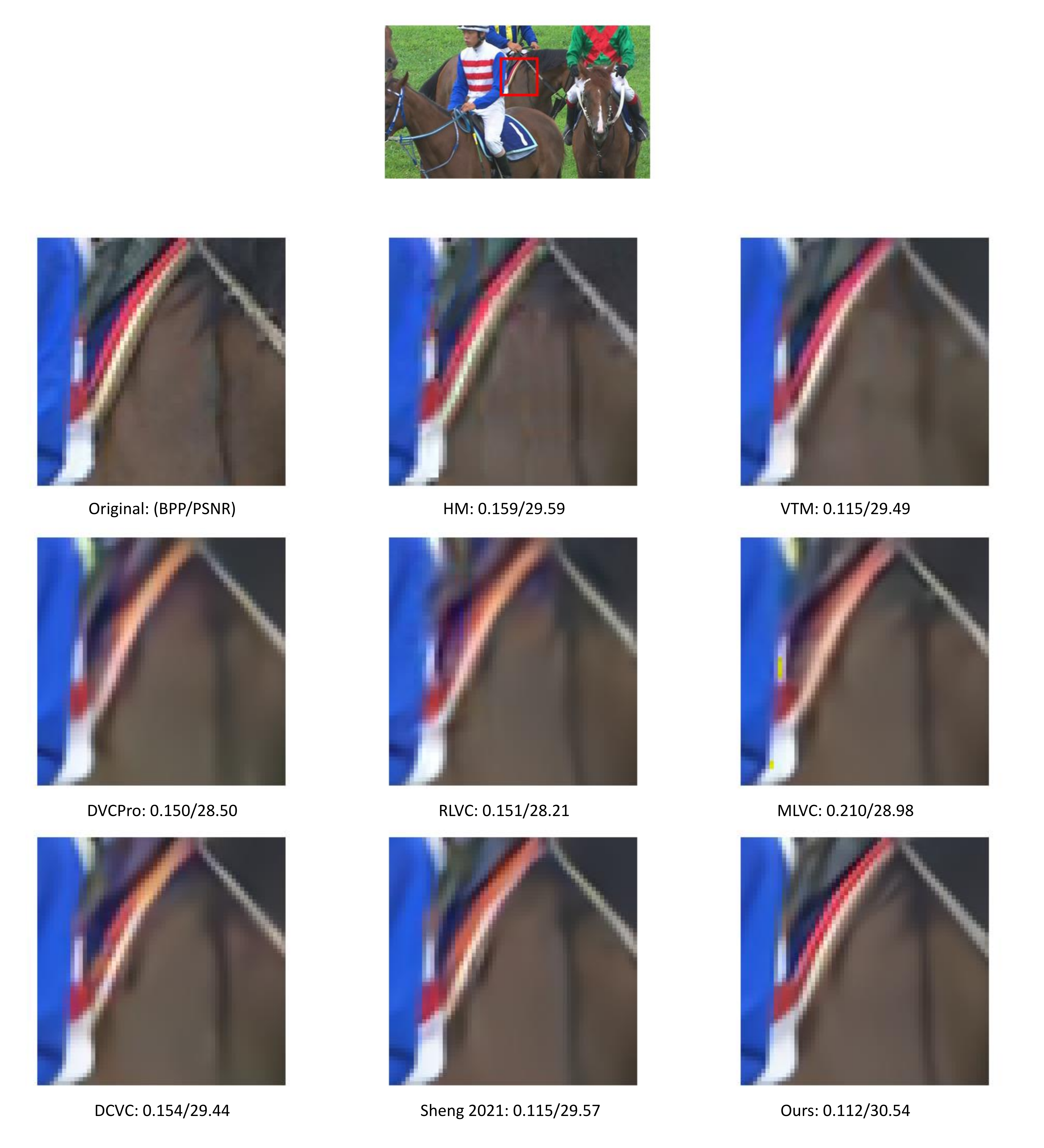} 
	\caption{ \textit{RaceHorses} from HEVC   D dataset. }
	\label{supp_visual1}
\end{figure*}

\begin{figure*}[t]
	\centering
	\includegraphics[width=1\linewidth]{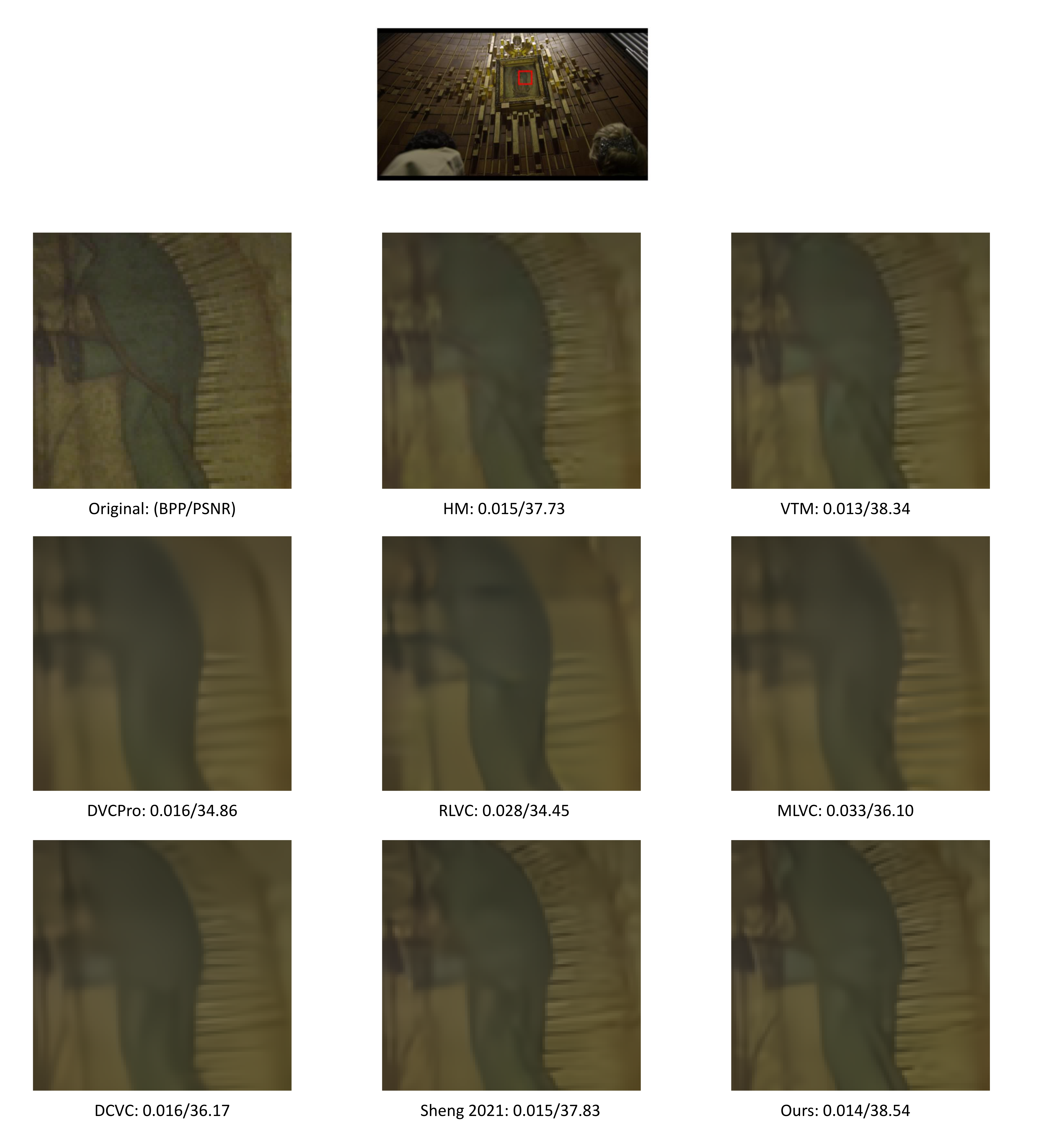} 
	\caption{ \textit{videoSRC03} from MCL-JCV dataset.  }
	\label{supp_visual2}
\end{figure*}

We also conduct the visual comparisons with the previous SOTA neural and traditional video codecs.  Two
examples are shown in Fig. \ref{supp_visual1} and Fig. \ref{supp_visual2}. From these comparisons, we can find that our codec can achieve much
higher reconstruction quality without increasing the bitrate cost. For instance, in the example shown
in Fig.  \ref{supp_visual1}, we can find that the frames reconstructed by previous neural codecs have obvious color
distortion. By contrast, our codec can restore more accurate   stripe color. In Fig. \ref{supp_visual2}, our codec can generate clearer texture details.

\section{Other Details}
Our codec is built upon previous developments DCVC \cite{li2021deep} and its improved work Sheng 2021 \cite{sheng2021temporal}. When compared with Sheng 2021 \cite{sheng2021temporal}, the motion estimation, motion vector encoder/decoder, temporal context mining, and contextual encoder/decoder are reused. The entropy model, quantization mechanism, and frame generator are redesigned. In addition, we add more details on our  quantizer and MEMC (motion estimation and motion compensation)  process, respectively.

\textbf{Quantizer}. The quantizer works as follows:

\begin{itemize}
	\item First, the latent representation is divided by the quantization step (QS).
	\item Second, the mean value is subtracted.
	\item Third, the latent representation is rounded to the nearest integer and converted into bit-stream via the arithmetic encoder.
\end{itemize}

During the decoding, the rounded latent representation is first decoded by the arithmetic decoder and adds back the mean value. At last, the QS is multiplied. In this process, most existing image and video codecs use the fixed QS, i.e., 1. By contrast, we decide the QS at multi-granularity levels. First, the global QS is set by the user for the specific target rate. Then it is scaled by the channel-wise QS because different channels contain information with different importance. At last, the spatial-channel-wise QS generated by our entropy model is applied. This can help our codec cope with various video contents and achieve precise rate adjustment at each position.

\textbf{Motion estimation and motion compensation}.  The MEMC process can be divided into the following steps:
\begin{itemize}
	\item First, the original motion vector (MV) between the current frame and the reference frame is estimated via the optical flow estimation network. The MV is dense at full resolution.
	\item The original MV is then transformed into latent representation via the MV encoder.
	\item The latent representation of MV is quantized and converted into bit-stream. The quantization step therein is also determined at multi-granularity levels. The quantization and entropy model are similar with those of the latent representation of the current frame.
	\item The MV at full resolution is then reconstructed via arithmetic decoder and MV decoder.
	\item The reconstructed MV is used to warp the feature and extract the context via the temporal context mining module.
\end{itemize}

\end{appendices}

\end{document}